\documentclass[twocolumn,
showpacs,preprintnumbers,
showkeys,
amsmath,amssymb,
prl,
superscriptaddress,aps,longbibliography
]{revtex4-2}
\usepackage{amsmath}
\usepackage{amsfonts}
\usepackage{amssymb}
\usepackage{changes}
\usepackage{placeins}

\renewcommand{\figurename}{Figure}
\newcommand{\Rnl}{\ensuremath{R_\mathrm{nl}}}
\newcommand{\Rcol}{\ensuremath{R_\mathrm{c}}}
\newcommand{\Rsq}{\ensuremath{R_\mathrm{sq}}}
\newcommand{\Vnl}{\ensuremath{V_\mathrm{nl}}}
\newcommand{\Vcnp}{\ensuremath{V_\mathrm{cnp}}}
\newcommand{\Vbg}{\ensuremath{V_\mathrm{bg}}}
\newcommand{\Vtg}{\ensuremath{V_\mathrm{tg}}}
\newcommand{\Cbg}{\ensuremath{C_\mathrm{bg}}}
\newcommand{\Ctg}{\ensuremath{C_\mathrm{tg}}}
\newcommand{\Bfocus}{\ensuremath{B_\mathrm{f}}}

\begin{document}
\title{Specular electron focusing between gate-defined quantum point contacts in bilayer graphene}

\author{Josep Ingla-Ayn\'es}
    \email[ ]{J.InglaAynes@tudelft.nl}
    \affiliation{Kavli Institute of Nanoscience, Delft University of Technology, Lorentzweg 1, 2628 CJ Delft, The Netherlands}
\author{Antonio L. R. Manesco}
    \affiliation{Kavli Institute of Nanoscience, Delft University of Technology, Lorentzweg 1, 2628 CJ Delft, The Netherlands}
\author{Talieh S. Ghiasi}
    \affiliation{Kavli Institute of Nanoscience, Delft University of Technology, Lorentzweg 1, 2628 CJ Delft, The Netherlands}
\author{Serhii Volosheniuk}
    \affiliation{Kavli Institute of Nanoscience, Delft University of Technology, Lorentzweg 1, 2628 CJ Delft, The Netherlands}
\author{Kenji Watanabe}
    \affiliation{Research Center for Functional Materials, National Institute for Materials Science, 1-1 Namiki, Tsukuba 305-0044, Japan}
\author{Takashi Taniguchi}
    \affiliation{International Center for Materials Nanoarchitectonics, National Institute for Materials Science,  1-1 Namiki, Tsukuba 305-0044, Japan}
\author{Herre S. J. van der Zant}
    \affiliation{Kavli Institute of Nanoscience, Delft University of Technology, Lorentzweg 1, 2628 CJ Delft, The Netherlands}
\date{\today} 

\begin{abstract}
We report on multiterminal measurements in a ballistic bilayer graphene (BLG) channel where multiple spin and valley-degenerate quantum point contacts (QPCs) are defined by electrostatic gating. By patterning QPCs of different shapes and along different crystallographic directions, we study the effect of size quantization and trigonal warping on the transverse electron focusing (TEF) spectra. 
Our TEF spectra show eight clear peaks with comparable amplitude and weak signatures of quantum interference at the lowest temperature, indicating that reflections at the gate-defined edges are specular and transport is phase coherent. 
The temperature dependence of the scattering rate indicates that electron-electron interactions play a dominant role in the charge relaxation process for electron doping and temperatures below 100~K. 
The achievement of specular reflection, which is expected to preserve the pseudospin information of the electron jets, is promising for the realization of ballistic interconnects for new valleytronic devices.
\end{abstract}


\maketitle

\section{Introduction}
Electronic devices with well-defined ballistic electron trajectories have triggered extensive research \cite{boggild2017,wang2019,lagasse2020, heinrich2021} and, to exploit their full potential, specular reflection of electron jets is a major requirement. Electrostatically-defined geometries are optimal platforms to realize the specular reflection, as shown by transverse electron focusing (TEF) measurements \cite{tsoi1974,tsoi1999,vanHouten1989,heremans1992,taychatanapat2013,lee2016,morikawa2015,bhandari2016,berdyugin2020,sonntag2020,bachmann2019}. 

 In this context, the exceptional electronic properties of graphene make it an ideal candidate for a wide variety of gate-defined devices where Klein tunneling enables new functionalities \cite{novoselov2007,cheianov2007,lee2015,chen2016, wang2019}. However, the absence of a bandgap complicates the creation of collimated beams and specular mirrors in graphene. The former has been realized by etching high-mobility graphene devices in absorptive pinhole collimators \cite{barnard2017}. The latter has been improved by recent fabrication progress, leading to the observation of multiple focusing peaks \cite{lee2016,berdyugin2020}. However, the reflection induced by disordered graphene edges is not specular \cite{walter2018}. This is a fundamental limitation that, in TEF experiments, results in a decrease of the peak amplitude as the number of reflections at the edge increases \cite{taychatanapat2013,lee2016,bhandari2016,berdyugin2020} and randomizes the valley degree of freedom \cite{walter2018}.
An alternative approach has been implemented in the quantum Hall regime, where the gaps between Landau levels have been used to create gate-defined interferometers \cite{wei2017,veyrat2019,ronen2021} and quantum point contacts (QPCs) \cite{ronen2021}. However, the effective confinement of carriers at $B=0$ in monolayer graphene remains a challenge.

In contrast, bilayer graphene (BLG) is a tunable-bandgap semiconductor with a trigonally-distorted Fermi surface \cite{castro2007,oostinga2008,zhang2009,icking2022}. It has recently been introduced as an ideal system for the realization of gate-defined QPCs \cite{allen2012,goossens2012,overweg2018,overweg2018PRL,kraft2018,velasco2018} capable of transmitting valley-polarized electron jets \cite{gold2021} and of hosting quantum dots with controllable spin and valley polarizations \cite{eich2018,banszerus2018}. 
Even though BLG hosts extraordinary properties, such as chirality-assisted cloaking \cite{gu2011,2lee2016} or anti-Klein tunneling \cite{katsnelson2006}, experiments on gate-defined BLG devices have so far focused on the characterization of QPCs \cite{allen2012,goossens2012,overweg2018,overweg2018PRL,kraft2018,velasco2018, gold2021}, quantum dots \cite{eich2018,banszerus2018,kurzmann2019}, quantum interference effects \cite{iwakiri2022}, and topological edge channels \cite{martin2008,san2009,li2016,lee2017,li2018}. 

In this work, we exploit the electrically-tuneable bandgap of BLG to create ballistic multiterminal BLG devices and measure TEF between gate-defined QPCs. We observe up to eight focusing peaks with comparable amplitudes, a clear indication of specular reflection at the gate-defined edges. Temperature-dependent measurements show that the TEF signal persists up to elevated temperatures and indicate the dominance of electron-electron interactions for electron doping and temperatures below 100~K.

\begin{figure}
	\centering
		\includegraphics[width=0.45\textwidth]{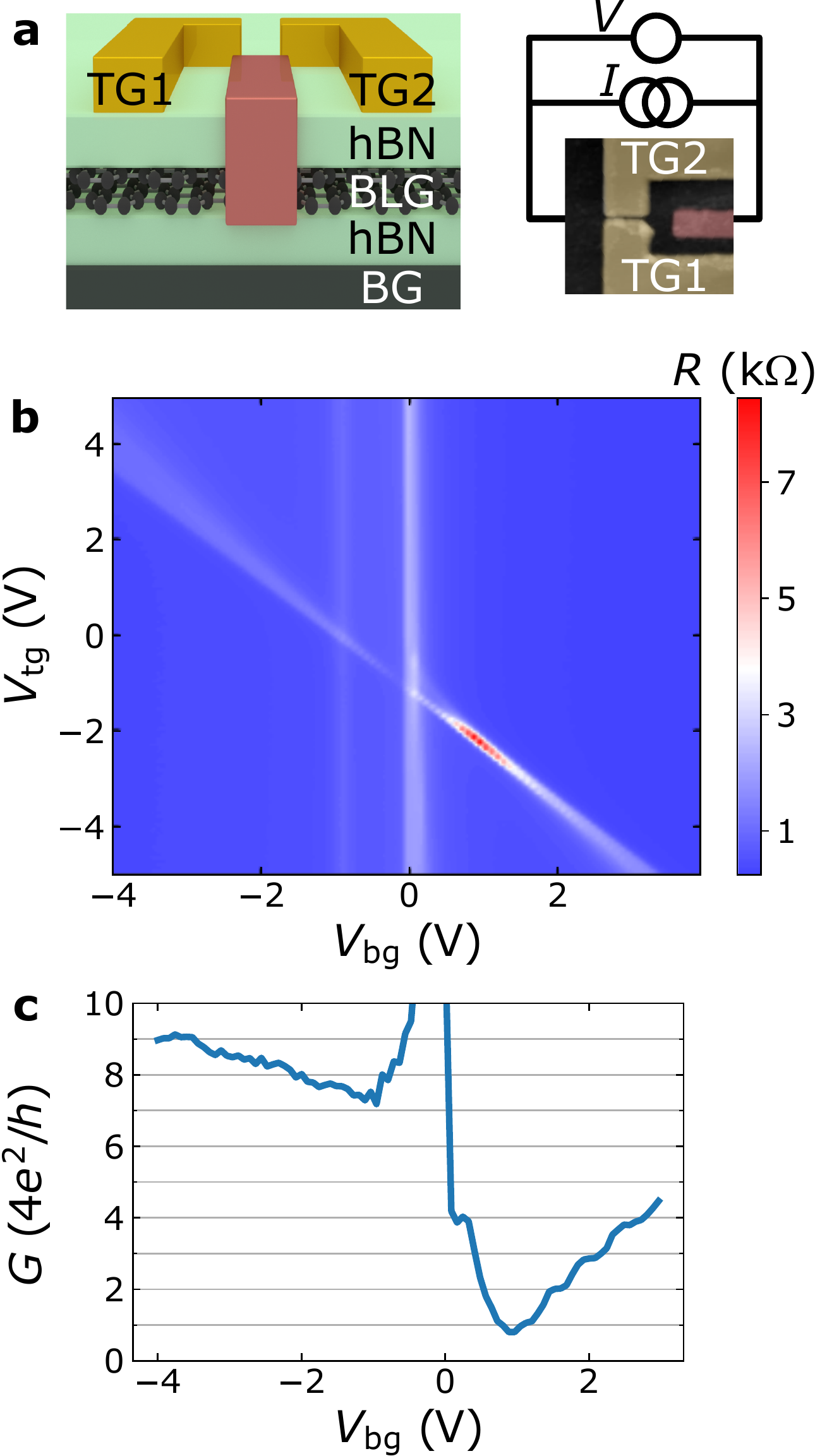}
	\caption{Gate-defined QPCs in BLG at 1.8~K. (a) Side (left) and top (right) view of the fabricated device. The top view is a false-color AFM image. The separation between the split top gates (TG1 and TG2) is approximately 50~nm and their width is 580~nm. At the side view (left panel), the hBN layers are green, the BLG and the few-layer graphene back gate (BG) are black. In both panels, the contacts to the BLG flake are brown and the top gates (TG) are dark yellow. (b) Two-terminal resistance ($R$) of one of the contacts used for the transverse electron focusing experiments as a function of \Vbg{} and \Vtg{}. \Vtg{} is the same for TG1 and TG2. (c) Point contact conductance obtained when the top-gated regions are charge neutral (see text for details).}
	\label{Figure1}
\end{figure}
\begin{figure*}
	\centering
		\includegraphics[width=\textwidth]{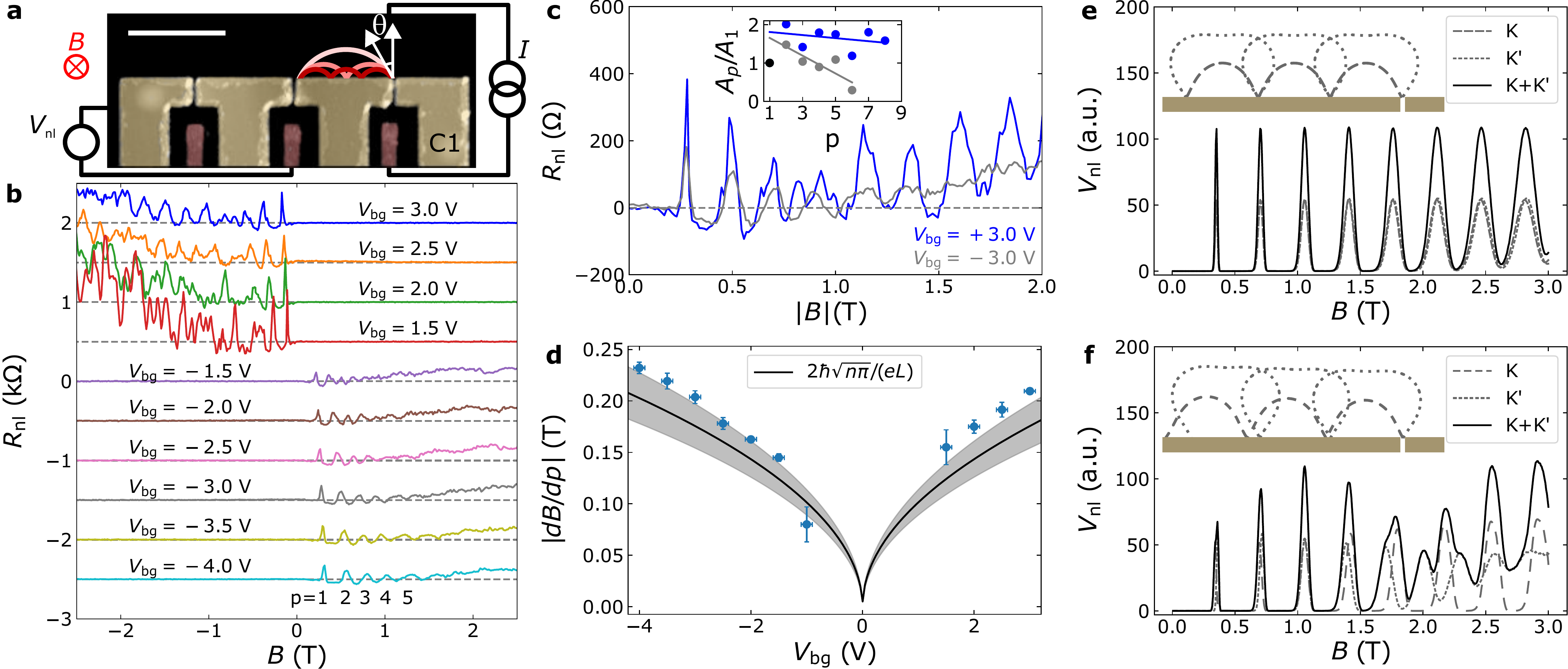}
	\caption{Transverse electron focusing between gate-defined QPCs in BLG at 1.8~K. (a) Measurement geometry. The nonlocal voltage (\Vnl{}) is measured as a function of $B$ while applying a current $I$ between the right QPC and a reference lead. The ballistic trajectories are sketched for the three first focusing peaks, which involve 0, 1 and 2 reflections with the gate-defined edge and assuming no trigonal warping. The scale bar is 2~$\mu$m. (b) Nonlocal resistance ($\Rnl{}=\Vnl{}/I$) as a function of $B$ for different \Vbg{} values. The dashed lines show the spectra offsets, which have been introduced for clarity. The offsets are shown by the dashed lines. \Vtg{} is tuned to follow the charge neutrality line of the top-gated regions (diagonal line in Fig.~\ref{Figure1}b). (c) Focusing spectra extracted from panel b at $\Vbg{}=\pm3$~V. A small offset in $B$ has been added to correct for the magnet remanence. The inset shows the evolution of the normalized area under the peaks ($A_p/A_1$) with $p$ (dots) and the lines are fits to illustrate the trends. (d) Peak separation as a function of \Vbg{}. The vertical error bars are the uncertainties from the \Bfocus{} vs.~$p$ linear fit  and the horizontal ones account for a 0.1~V uncertainty of the CNP. The black line is the result from Eq.~\ref{Equation1} assuming normal incidence from the QPCs ($\theta=0$). The gray area corresponds to the experimental error from determining $n$ (14\%) and $L$ (10\%). Simulated TEF signal for perfectly aligned (e) and 3$^\circ$ misaligned (f) QPCs with respect to the armchair crystallographic direction. The black curves have been obtained by adding the $K$ and $K'$ valley-resolved spectra. The insets show the trigonally-warped trajectories corresponding to the average incidence angles for valleys $K$ and $K'$ and the dark yellow rectangles represent the gate-defined edges.}
	\label{Figure2}
\end{figure*}
\section{Results}

We fabricated two double-gated, boron nitride (hBN)-encapsulated BLG heterostructures on few-layer graphene back gates, each containing multiple devices using the dry transfer technique described in \cite{zomer2014, purdie2018}. The electrodes were defined using conventional e-beam lithography. The BLG flakes were connected to Ti/Au electrodes (brown rectangles in Fig.~\ref{Figure1}a) after using a CHF$_3$/O$_2$ plasma to etch the upper hBN and BLG layers at the contact area \cite{wang2013}. The top gates, which are dark yellow in Fig.~\ref{Figure1}a, were deposited on the top hBN (see SI section S1 for the fabrication details). 
The side and top view images of a typical QPC are shown in Fig.~\ref{Figure1}a. 
Here we discuss the results on the first heterostructure (Sample~1); the results on Sample~2 are shown in the SI section~S9. 

The two-terminal resistance of the QPC, defined as $R = V/I$, where $V$ and $I$ are the measured voltage and applied current, respectively (see Fig.~\ref{Figure1}a, right panel) has been recorded as a function of the top gate voltage (\Vtg{}) and the back gate voltage (\Vbg{}). As shown in Fig.~\ref{Figure1}b, three features can be distinguished from this result: The first one is a vertical line at $\Vbg{}\approx 0$, which corresponds to the charge neutrality point (CNP) of the non-top-gated BLG channel. The CNP does not occur at exactly $\Vtg{}=0$ due to a small hole-doping. The second feature is a faint vertical line at $\Vbg{}\approx-1$~V. Four-terminal measurements (see SI section~S3) indicate that it corresponds to the CNP of the BLG near the Ti/Au contacts, where the top hBN and BLG have been etched. 

The last feature is a diagonal line that has a negative slope (\Vtg{} decreases as \Vbg{} increases) that corresponds to the CNP of the regions under TG1 and TG2. Since both \Vbg{} and \Vtg{} influence the carrier density ($n$) at these regions, the introduction of electrons by \Vbg{} to the BLG channel must be counteracted by an opposite \Vtg{} to keep the channel charge neutral. We use the slope of this line to obtain the ratio between the top gate (\Ctg{}) and back gate (\Cbg{}) capacitances: $\Cbg{}/\Ctg{}=-\Delta \Vtg{}/\Delta \Vbg{}\approx 1.22$. This value is consistent with the factor 1.22 obtained from the ratio between the hBN-flake thicknesses extracted from AFM imaging (see SI section~S1).
Even though the electric field applied by the gates opens a bandgap in the double-gated BLG regions which increases with $|\Vbg{}|$ \cite{castro2007,oostinga2008,zhang2009,icking2022}, the resistance along the diagonal line does not increase with $|\Vbg{}|$. This is due to the small gap between TG1 and TG2 (Fig.~\ref{Figure1}a). In this region the carrier density is not zero, leading to the formation of a \Vbg{}-controlled QPC with tuneable carrier density.

To determine if the QPC conductance ($G$) is quantized, we have determined its resistance by taking, for each \Vbg{}, the difference between the maximal and minimal $R$. This operation allows us to subtract the resistance of the Ti/Au contacts and the BLG regions that are not affected by \Vtg{}. The result is shown in Fig.~\ref{Figure1}c. For negative \Vbg{}, $G$ shows values higher than $7\times4e^2/h$ and it changes in a monotonic way with small oscillations. In contrast, for positive \Vbg{}, $G$ shows four steps at $G=N\times 4e^2/h$ with $N=1, 2, 3,~\mathrm{and}\,4$. This behavior, which is reproduced in five of the six QPCs characterized, indicates the formation of a spin and valley-degenerate QPC \cite{vanWees1988, overweg2018,overweg2018PRL, kraft2018}. Note that the sharp increase of $G$ near $\Vbg{}=0$ is a consequence of the extraction method when there is no bandgap under the double-gated regions and $R$ shows very small changes with \Vtg{}.
Even though the reason for the electron-hole asymmetry is not clear, we believe that one possibility may be a residual doping of the double-gated regions caused by the fabrication. Since the QPC region is not affected by this process, the potential landscape could become asymmetric to the sign reversal of the gate voltages. This could make the QPC narrower for electron than hole doping or modify its carrier density.

 When a magnetic field ($B$) is applied perpendicular to the plane of a ballistic BLG device, electrons deviate from their straight trajectories by the Lorentz force. If the Fermi surface is circular, they follow circular orbits with radius $r_c=\hbar k_F/eB$, where $\hbar$ is the reduced Plank constant and $k_F$ is the Fermi wavevector ($k_F=\sqrt{n\pi}$). As a consequence, the transmission between different contacts connected at a distance $L$ from each other shows maxima at magnetic fields (\Bfocus{}) given by \cite{tsoi1974,vanHouten1989} 
\begin{equation}
\Bfocus{}=\frac{2p\hbar k_F\cos{\theta}}{eL},
\label{Equation1}
\end{equation}
where $\theta$ is the angle at which the electron flow departs from the emitter, $L=2\,\mu$m is the injector-detector distance, and $p = 1, 2, 3,\cdots,n$ is an integer which accounts for the $p-1$ reflections that occur at the device edge between the contacts (Fig.~\ref{Figure2}a). 

\begin{figure}
	\centering
		\includegraphics[width=0.45\textwidth]{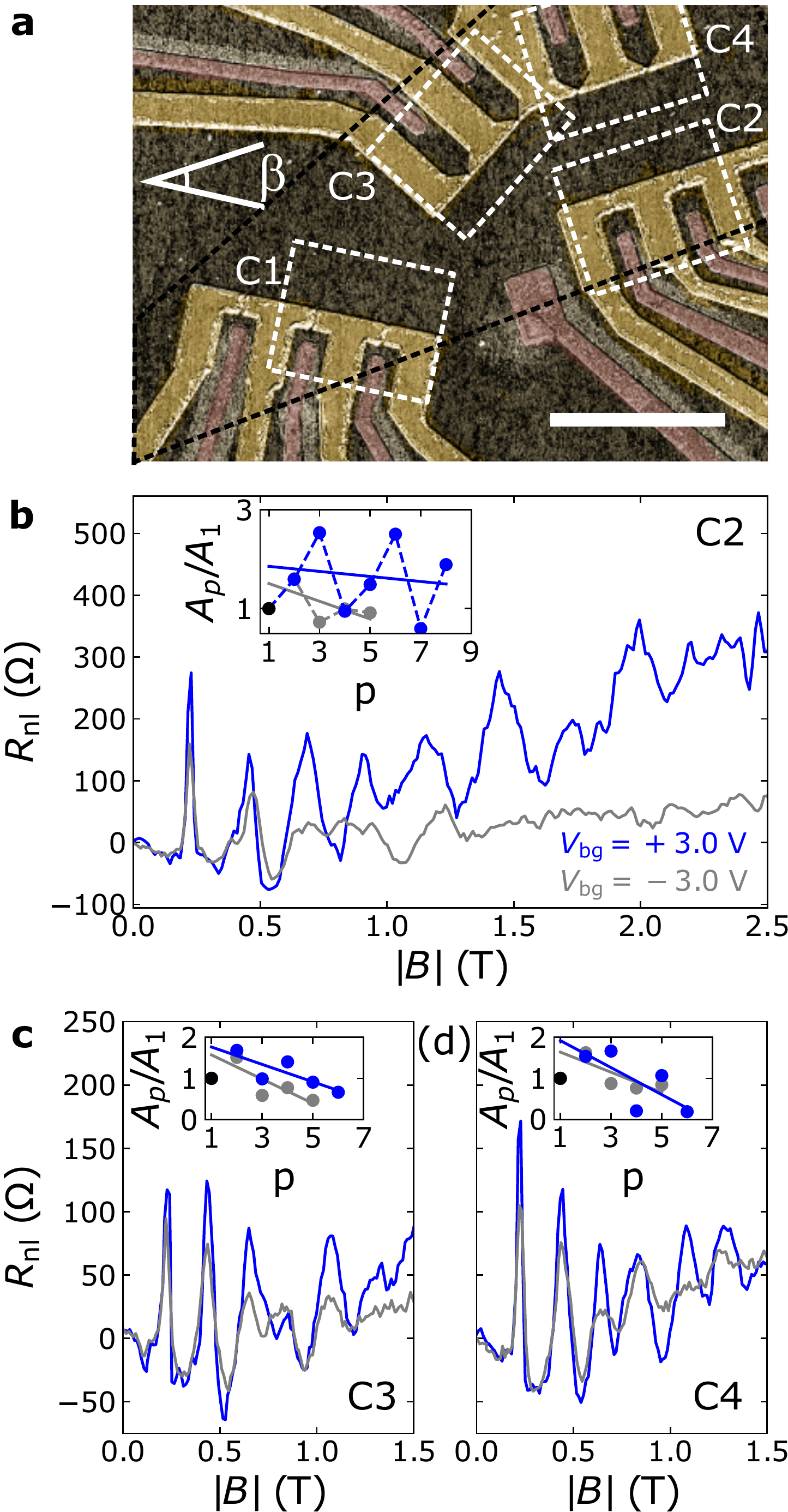}
	\caption{TEF along different crystallographic directions at 1.8~K. (a) False-color AFM image with the QPCs involved in C1 (Fig.~\ref{Figure2}), C2, C3 and C4. C2 is rotated an angle $\beta=30^\circ$ with respect to C1 and C4 is rotated $\beta=-30^\circ$ with respect to C3. The scale bar is 5~$\mu$m. (b-d) TEF in configurations C2-C4 at \Vbg{}$=\pm3$~V. The insets show the normalized peak area vs $p$ and the lines are linear fits to illustrate the trend.}
	\label{Figure3}
\end{figure}

TEF measurements have been performed using configuration C1, which is shown in Fig.~\ref{Figure2}a. A current ($I$) is applied to the right QPC to generate an electron flow into the ballistic BLG channel that is steered using the out-of-plane $B$-field. To detect the ballistic skipping orbits, the nonlocal voltage (\Vnl{}) is measured between the left QPC and a reference electrode connected further at the left of the BLG channel. To avoid voltage \Vnl{} offsets, we have used a differential DC measurement technique to obtain the TEF in Figs.~\ref{Figure2} and \ref{Figure3}.

The results from such measurements performed for different \Vbg{} are shown in Fig.~\ref{Figure2}b. Note that, to assure that the charge transport occurs only through the QPCs, we have adjusted \Vtg{} to keep the double-gated regions charge neutral (diagonal line in Fig.~\ref{Figure1}b). 
We first consider the $\Vbg{}=-4$~V case. For $B<0$, the signal is zero (dashed lines) or smaller than the noise level of the measurement, which is $2~\Omega$, consistent with the fact that the ballistic electron stream deviates towards the right and does not generate a signal on the detector. In contrast, when $B>0$, five clear focusing peaks are observed, indicating that even though the QPC conductance is not quantized for $\Vbg{}<0$ (Fig.~\ref{Figure1}d), the hole trajectories are well-defined and reflection at the gate-defined edge between both QPCs is smooth. 
As \Vbg{} approaches zero, $n$ in the BLG channel decreases and the distance between the peaks becomes smaller.
At $\Vbg{}>0$ peaks occur for $B<0$, consistent with the fact that the carriers have changed from holes to electrons \cite{taychatanapat2013, lee2016,berdyugin2020}.

\begin{figure}
	\centering
		\includegraphics[width=0.4\textwidth]{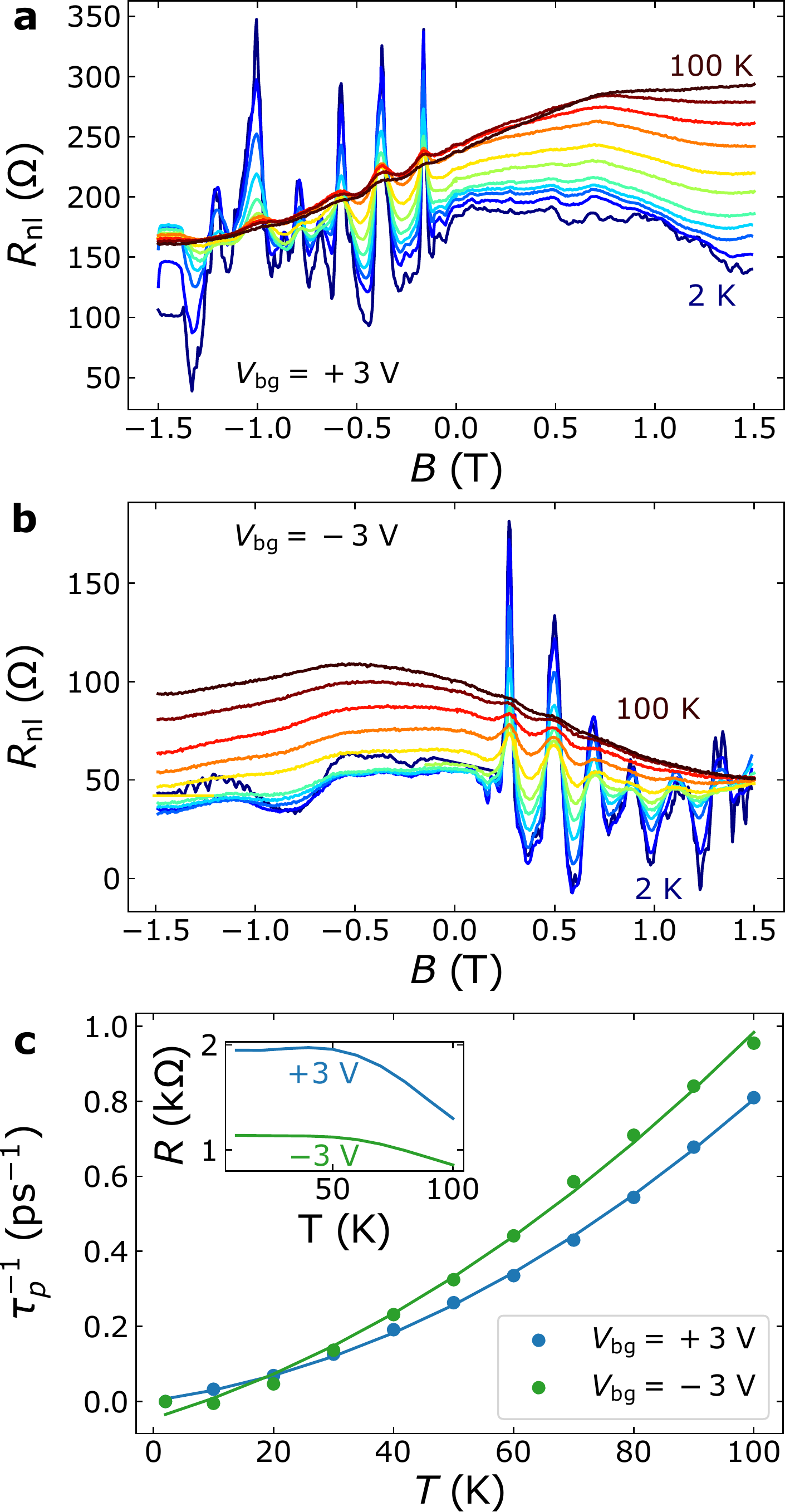}
	\caption{Temperature dependence of TEF in BLG. $B$-dependence of \Rnl{} for $T=2,\,10,\, 20,\,...,\,100$~K at (a) $\Vbg{}=+3$~V and (b) $\Vbg{}=-3$~V. (c) Scattering rate estimated using Equation~\ref{EqnTp} (dots) and its fit to a parabola (lines). The inset shows the $T$-dependence of the QPC resistance.}
	\label{Figure4}
\end{figure}

For a more detailed comparison, in Fig.~\ref{Figure2}c we show the $\Vbg{}=\pm3$~V spectra.  
Two clear differences can be distinguished: i) The $p=1$ peak is two times higher for $\Vbg{}=+3$~V. This is most likely due to the lower $G$ at $\Vbg{}=+3$~V, which converts the collector current ($I_c$) into the measured \Vnl{}$=I_c/G$. As shown in Fig.~\ref{Figure1}c, $G$ is roughly two times larger for $\Vbg{}=-3$~V than for $\Vbg{}=+3$~V, explaining most of the measured asymmetry in the $p=1$ peak magnitude. ii) The peak amplitude decays with $p$ much faster at $\Vbg{}=-3$~V.
To quantify the TEF signal decay with $p$ and correct for a small contact magnetoresistance (see SI section~S4), we calculated the area under the TEF peaks \cite{lee2016} normalized by the two-terminal resistance (see SI section~S5b). The result is shown in the inset of Fig.~\ref{Figure2}c with a linear fit excluding the $p=1$ peak (which has the smallest area). The obtained peak areas are fairly constant from $p=2$ up to $p=8$ (including the $p=4$ peak, that occurs between 0.75 and 1~T and is split in two), indicating specular reflection. In contrast, for \Vbg{}$=-3$~V, the peak area decays with increasing $p$.
 
The faster peak decay for \Vbg{}$<0$ can be explained in terms of a change of the QPC width ($W$). 
The finite $W$ of the detector poses an upper bound to the maximum number of peaks that can be measured. In particular, if $r_c\cos{\theta}\leq W/2$, all electrons will enter the detector and extra peaks cannot be detected \cite{vanHouten1989}, leading to $B\leq 2$~T for $W=200$~nm, $\cos{\theta}=1$ and a circular trajectory. In contrast, for $W=100$~nm, we obtain $B\leq 3.9$~T. 
As shown in Fig.~\ref{Figure1}c, $G$ is almost eight times smaller for electrons than for holes, indicating that a significant electron-hole asymmetry in the QPC width is plausible.
Additionally, decreasing the injector $W$ is known to lead to electron jets with improved collimation \cite{molenkamp1990,barnard2017}. Since the focusing length of a trajectory depends on its injection angle, the differences between focusing lengths of different trajectories increase with $p$. Thus, a narrow angular distribution is expected to help maintaining a constant peak amplitude, even after several edge reflections. 

In Fig~\ref{Figure2}b, for $\Vbg{}=1.5$~V (red curve), additional oscillations similar to those in Refs.~\cite{vanHouten1989,ronen2021} can be observed on top of the focusing spectrum. The amplitude of these oscillations decreases with increasing \Vbg{}, a result which is consistent with quantum interference between the different electron paths contributing to the TEF signal because the Fermi wavelength increases with decreasing $n$.

For completeness, we also measured \Rnl{} near $\Vbg{}=0$~V, where the double-gated regions are not gapped. In this case, we observe large background signals and clear plateaus, indicating that there is a significant current leakage through the top-gated regions (see SI section~S7).

To gain more insight into the measured TEF spectra, we have analyzed the positions of the focusing peaks (\Bfocus{}) as a function of $p$. In particular, we determined \Bfocus{} and fit it to $\Bfocus{}=B_0+(dB/dp)\times p$, where $B_0$ and $dB/dp$ are constants accounting for the magnet remanence and the average spacing between the peaks, respectively. In Fig.~\ref{Figure2}d we show $|dB/dp|$ and compare it with the result from Eq.~\ref{Equation1} for normal incidence ($\theta=0$). The agreement between both curves further confirms that our signal is due to TEF.

The results shown in Fig.~\ref{Figure2}c at $\Vbg{}=+3$~V show features resembling a beating pattern. In particular, all the peaks except $p=1$ and $4$ can be decomposed into two narrower peaks and the latter, which has a dip where one would expect a peak, can be decomposed into three well-separated peaks. Additionally, the Fourier transform of the TEF spectrum (see SI section~S5b for details) also indicates the presence of a beating pattern, implying a periodic modulation.
Even though there may be a combination of impurities that could explain this effect, there is a fundamental reason to expect such features in the TEF spectra. BLG is known for showing trigonal warping, i.e., its Fermi surface is not circular. In this case, the emission of electrons by the QPCs occurs in jets that depend on the crystallographic orientation of the QPCs on the BLG \cite{gold2021, manesco2022}. If the QPCs are slightly misaligned with respect to a crystallographic direction, the valley-polarized jets will be emitted with slightly different $|\theta|$, leading to two different \Bfocus{} for the peaks in valleys $K$ and $K'$. 

To determine whether this scenario is compatible with the TEF spectrum in Fig.~\ref{Figure2}c, we have performed semiclassical calculations considering the effect of trigonal warping on the electronic trajectories and their angular distribution (see SI section~S9 for details). The results are shown in Figs.~\ref{Figure2}e and \ref{Figure2}f for the perfectly aligned and the small misalignment (0.05~rad~$\approx~3^\circ$) cases, respectively. The trajectories are shown in the insets. In the latter, a beating pattern arises which is compatible with the measured data.

To show the robustness of the TEF measurements and explore the role of the BLG crystallographic orientation on the TEF spectra, we have patterned QPCs in different directions on the same BLG flake. The relative angle between the QPC sets is $30^\circ$ to compare the armchair with the zigzag crystallographic directions. Since C2 is aligned parallel to the longest BLG straight edge (black dashed line in Fig.~\ref{Figure3}a), we expect the C2 QPCs to be aligned with a crystallographic direction \cite{you2008}. Thus, the 30$^\circ$ rotated C1 QPCs, are expected to be along the other.
We compare the TEF spectra in Fig.~\ref{Figure2}c with the TEF spectra obtained using configurations C2, C3, and C4 from Fig.~\ref{Figure3}a for $\Vbg{}=\pm3$~V, shown in Figs.~\ref{Figure3}b, \ref{Figure3}c and \ref{Figure3}d, respectively. The results show several features: i) The TEF peaks decay faster with $p$ for holes than for electrons in all the geometries. ii) For C3 and C4, which contain horn-like QPCs not showing size quantization (see SI section~S6 for details), the decay in peak amplitude for electrons is more pronounced than for C1 and C2 where $G$ is quantized. As a consequence, six peaks can be distinguished instead of eight. iii) The width of the $p=1$ peak is significantly smaller than that of the $p=2$ peak in all the configurations, both for electron and hole doping.
Observations i) and ii) show a correlation between $G$ and the TEF peak amplitude decay, further indicating that the QPC width plays a relevant role in the peak amplitude decrease.

It is worth noting that the spectrum in Fig.~\ref{Figure3}b at $\Vbg{}=+3$~V (using QPCs with quantized conductance) does not show a beating pattern as in Fig~\ref{Figure2}c. As shown in Figs.~\ref{Figure2}e and \ref{Figure2}f, the occurrence of a beating pattern is very sensitive to a tiny misalignment. Thus, the absence of such a pattern in C2 is consistent with the Fermi surface having some degree of trigonal warping.  

Finally, from the comparison between C1 and C2, which are aligned along different crystallographic directions on the same BLG flake, one would expect that, in one of the configurations, one of the valleys ($K$) emits an electron jet with $\theta=0$ and the other valley ($K'$) emits two jets at $\pm60^\circ$. As a consequence, \Bfocus{} for the electrons in $K'$ is approximately half the \Bfocus{} in valley $K$. As a consequence, an even-odd effect arises where the even $p$ peaks are twice as large as the odd $p$ ones.
The periodic modulation in $A_p/A_1$ shown at the inset of Fig.~\ref{Figure3}b may be a signature of such effect, but the difficulty determining the background level (see SI section~S5), the absence of a clear modulation of the peak heights and the different widths of the TEF peaks challenge such interpretation.

To characterize the scattering sources in BLG, we have measured \Rnl{} vs.~$B$ at different temperatures ($T$) at $\Vbg{}=\pm3$~V. At 2~K, the peak height is the highest, and, as $T$ increases, the background becomes more pronounced and the focusing signal gets smaller. 
Comparing the 2~K with the 10~K measurements, the 2~K spectra contain extra features at positive and negative $B$-fields. A fast decay when increasing $T$ indicates that these features are likely due to quantum interference, as the phase-coherence length is known to drop within this range \cite{kozikov2012}.

To extract the $T$-dependence of the scattering rate ($\tau_p^{-1}$) from Figs.~\ref{Figure4}a and \ref{Figure4}b, we have used \cite{lee2016}:
\begin{equation}
    \tau_p^{-1}=-2v_\mathrm{F}/(\pi L)\log(A_2(T)/A_2(T_\mathrm{base})),
    \label{EqnTp}
\end{equation}
where $v_\mathrm{F}$ is the Fermi velocity, $A_2(T)$ the area under the second focusing peak at each $T$, and $A_2(T_\mathrm{base})$ the area of the second peak at $T=2$~K (see SI section~S5 for the results obtained using the area under the first peak). 
As shown in the inset of Fig.~\ref{Figure4}c, $R$ decreases significantly above 50~K, most likely due to thermal activation of the double-gated BLG regions \cite{icking2022}. To take into account the $T$-dependent $R$, we have normalized \Rnl{} by $R$ to obtain the area under each peak. The result for $p=2$ is shown in Fig.~\ref{Figure4}c. Here, the dots correspond to the values extracted from Figs.~\ref{Figure4}a and \ref{Figure4}b, and the solid lines are fits to parabolas ($\tau_p^{-1}=aT^2+bT+c$). A quadratic $T$-dependence of $\tau_p^{-1}$ is associated with electron-electron interactions \cite{lee2016,bandurin2016,bandurin2018}. In contrast, a linear dependence is associated with phonon-dominated scattering \cite{taychatanapat2013,hwang2008}. By calculating $T_0=b/a$, which is the $T$ where the quadratic term starts to dominate over the linear term, we obtain $T_0\approx 40$ $(90)$~K for $\Vbg{}=+(-)3$~V, indicating that 
 electron-electron interactions play a relevant role in the $T$-dependent scattering for electrons, but not for holes, see SI section~S5 for the fitting parameters and a more detailed discussion. 
 Note that the results shown in Figs.~\ref{Figure4}a and \ref{Figure4}b were obtained in a different cooldown and using a lock-in technique. We suspect that a slight miscalibration of \Vtg{} has led to larger background signals than in the previous measurements.

\section{Conclusions}
To conclude, we have measured TEF in hBN-encapsulated BLG devices where QPCs are defined in different directions using electrostatic gating. Our results show eight focusing peaks with similar amplitude together with quantum interference features. By comparing TEF spectra with semiclassic simulations we identify a periodic modulation of the peak size that is consistent with the effect of trigonal warping.  
Moreover, the TEF temperature dependence shows that the signal persist up to 100~K and indicates that, for positive \Vbg{}, electron-electron interactions play an important role in the charge relaxation process at elevated temperatures. Our results are promising for future valleytronic devices.

\section{Data availability} 
All the data and code associated with the analysis and theoretical simulations are available free of charge at \cite{zenodo}.

\section{Acknowledgements}
 We thank Prof.~K.~Ensslin and K.~Vilkelis for insightful discussions. This project received funding from the European Union Horizon 2020 research and innovation program under grant agreement no. 863098 (SPRING). JI-A acknowledges support from the European Commission for a Marie Sklodowska–Curie individual fellowship No. 101027187-PCSV. ALRM work was supported by VIDI grant 016.Vidi.189.180. K.W. and T.T. acknowledge support from JSPS KAKENHI (Grant Numbers 19H05790, 20H00354 and 21H05233).

\bibliography{bibliography}


\pagebreak
\widetext
\begin{center}
\textbf{\large Supplementary information of ``Specular electron focusing between gate-defined quantum point contacts in bilayer graphene"}
\end{center}
\renewcommand{\thetable}{S\arabic{table}}
\renewcommand{\theequation}{S\arabic{equation}}
\renewcommand{\figurename}{Figure}
\renewcommand{\thefigure}{S\arabic{figure}}
\renewcommand\thesection{S\arabic{section}}
\setcounter{equation}{0}
\setcounter{figure}{0}
\setcounter{table}{0}
\setcounter{page}{1}
\makeatletter

\tableofcontents
\section{Device fabrication}
\begin{figure}
\centering
\includegraphics[width=\textwidth]{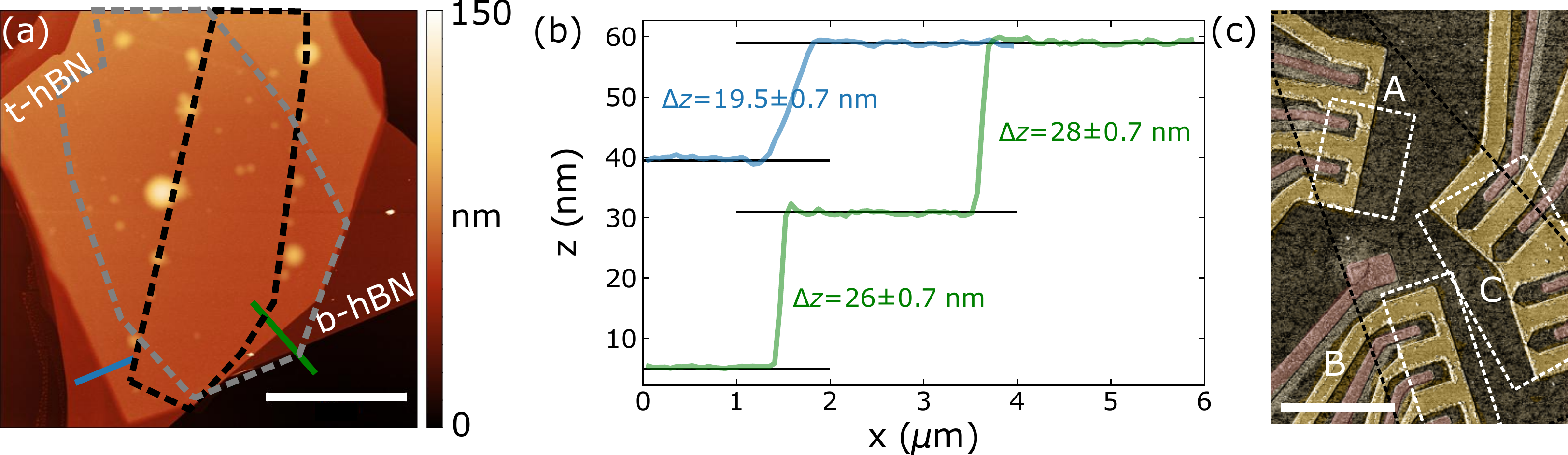}
\caption{(a) AFM image of the device after annealing. The black dashed edge corresponds to the BLG, the gray edge to the multilayer graphene backgate and t(b)-hBN to the top (bottom) hBN flakes. The scale bar is 10~$\mu$m. (b) Height profiles extracted along the blue and green lines in panel a with the extracted step heights corresponding to the hBN thicknesses. The error range accounts for the roughness of the hBN surfaces (0.5~nm). (c) Phase of an AFM image of the completed device with the BLG edges in black. The dark yellow Ti/Au structures are top gates used to define QPCs and the brown ones are contacts to the BLG. The scale bar is 5~$\mu$m. \label{FigureAFM}}
\end{figure}

Samples have been prepared using the poly (bisphenol A) carbonate (PC) technique \cite{zomer2014, purdie2018}. The hexagonal boron nitride (hBN) and bilayer graphene (BLG) flakes were exfoliated from bulk crystals \cite{novoselov2004} and picked up with a PC layer at temperatures between 60 and 90~$^\circ$C. The resulting heterostructure was released on a clean SiO$_2$ substrate with Au markers by melting the PC layer at temperatures above 150~$^\circ$C. Subsequently, the PC covering the stack  was removed from the surface by dissolving it in chloroform. The stack was then annealed for 1~h in an Ar atmosphere at 400~$^\circ$C before contact preparation. At this stage, an atomic force microscopy (AFM) image was taken in the AC mode of a Cypher AFM to determine the thickness of the different flakes (Fig.~\ref{FigureAFM}a). The profiles extracted to obtain the thickness of the hBN flakes are indicated by a green and a blue line and shown in Fig.~\ref{FigureAFM}b, together with the estimated flake thicknesses. The bottom hBN thickness with respect to the SiO$_2$ (left step green profile) is 26~nm, 6.5~nm larger than the 19.5~nm obtained when the bottom hBN is covered by the top hBN (blue profile). On the one hand, this discrepancy may be due to the different adhesion of SiO$_2$ and hBN surfaces, affecting the left step of the green profile. On the other hand, the slight negative slope of the lower plateau of the blue line, combined with its smoother profile (caused by the top hBN coverage) may lead to a thickness underestimation. We take the average (23~nm) as the estimate of the bottom hBN thickness.  

The contacts to the bilayer graphene were defined using e-beam lithography. After defining the contact pattern, the top hBN was etched with a mixture of CHF$_3$ and O$_2$ with a 10 to 1 flow ratio, 40~W of power and a pressure of 5~$\mu$bar \cite{wang2013}. This recipe gives an etch rate for hBN of approximately 30~nm/min and leaves the BLG edges exposed. After etching, the Ti(5~nm)/Au(35~nm) electrodes were deposited using e-beam evaporation. The top gates were prepared using the same method replacing the CHF$_3$/O$_2$ etching by a mild O$_2$ etching (10~s 15~W) to promote adhesion between the Ti and the hBN surfaces.
The final result is shown in Fig.~\ref{FigureAFM}c: Rectangle A surrounds the contacts involved in C1, rectangle B the contacts in C2 and C2L, and rectangle C, the contacts in C3 and C4 (see Section~\ref{SectionTEFGeoms}). Note, that Fig.~\ref{FigureAFM}c corresponds to the false-colored phase channel of an AFM image. The contacts to the BLG (dark dashed line) are red and the top gates (yellow) surround the contacts to create the electrostatically-defined quantum point contacts (QPCs). There is one extra contact above and another one below the AFM image which were used as reference for the nonlocal measurements.
\FloatBarrier
\section{Backgate capacitance obtained using Shubnikov–de Haas oscillations}
\begin{figure}
\centering
\includegraphics[width=\textwidth]{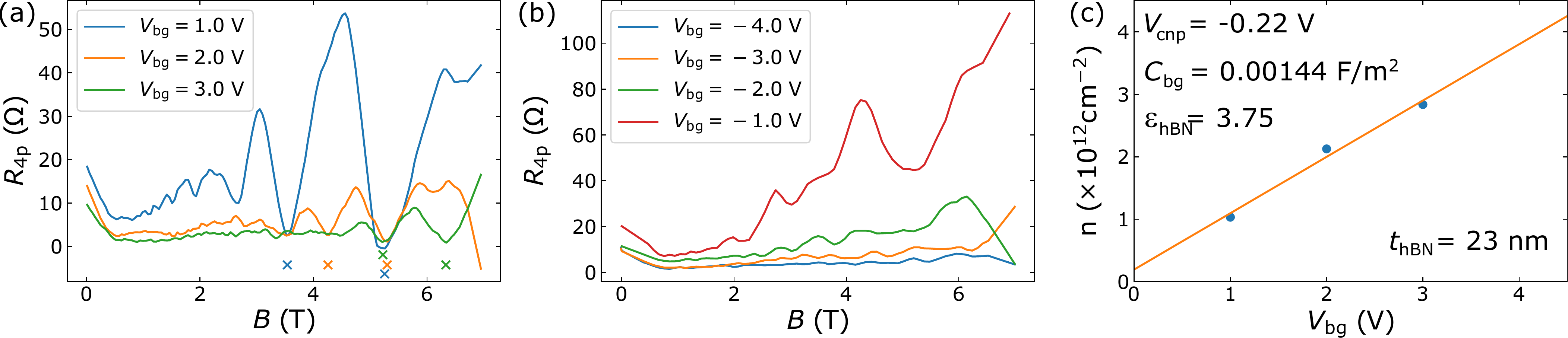}
\caption{(a) and (b) $R_{4p}$ vs.~$B$ up to 7~T for positive and negative \Vbg{}, respectively. The crosses in panel a indicate the $B$-values used to obtain the carrier density. (c) Carrier density extracted from panel a at \Vbg{}$=$1, 2 and 3~V with the linear fit to the $n$ vs.~\Vbg{} dependence used to extract the backgate capacitance.\label{SdHOscillations}}
\end{figure}

To determine the capacitance of the backgate (\Cbg{}) in an accurate way we have used Shubnikov-de Haas oscillations. In a four-terminal measurement configuration, we measured the longitudinal four-point resistance ($R_{4p}$) at different \Vbg{} while sweeping $B$ up to 7~T. The results from such a measurement are shown in Figs.~\ref{SdHOscillations}a and \ref{SdHOscillations}b for positive and negative \Vbg{}, respectively. Since for negative \Vbg{} the measured data does not show clear oscillations to extract the carrier density ($n$), we have only used positive \Vbg{}.
To determine $n$, we have used \cite{novoselov2004}:
\begin{equation}
n=\frac{4 e}{h}\frac{B_{i+1}B_i}{B_{i+1}-B_{i}}
\label{EqNSdH}
\end{equation}
where $e$ is the electron charge, $h$ is the Plank constant and $B_i$ and $B_{i+1}$ correspond to the adjacent field positions where $R_{4p}$ is minimal, indicated as crosses in Fig.~\ref{SdHOscillations}a. The result from Equation~\ref{EqNSdH} is shown in Fig.~\ref{SdHOscillations}c and plotted vs.~\Vbg{} as blue dots. Because the carrier density changes with the gate voltage following $n=\Cbg{}(\Vbg{}-\Vcnp{})/e$, where \Vcnp{} is the position of the charge neutrality point, we have used a linear fit $n=A\Vbg{}+B$ (orange line) to extract \Cbg{}$=Ae$ and $\Vcnp{}=-B/A$. With $\epsilon_\mathrm{hBN}=\Cbg{}t_\mathrm{hBN}/\epsilon_0$, where $\epsilon_0$ is the vacuum permittivity, and the bottom hBN thickness from AFM measurements ($t_\mathrm{hBN}\approx23$~nm), we estimate its dielectric constant, $\epsilon_\mathrm{hBN}\approx3.75$, in agreement with \cite{laturia2018}. 

\section{Electronic mobility}
\begin{figure}
\centering
\includegraphics[width=\textwidth]{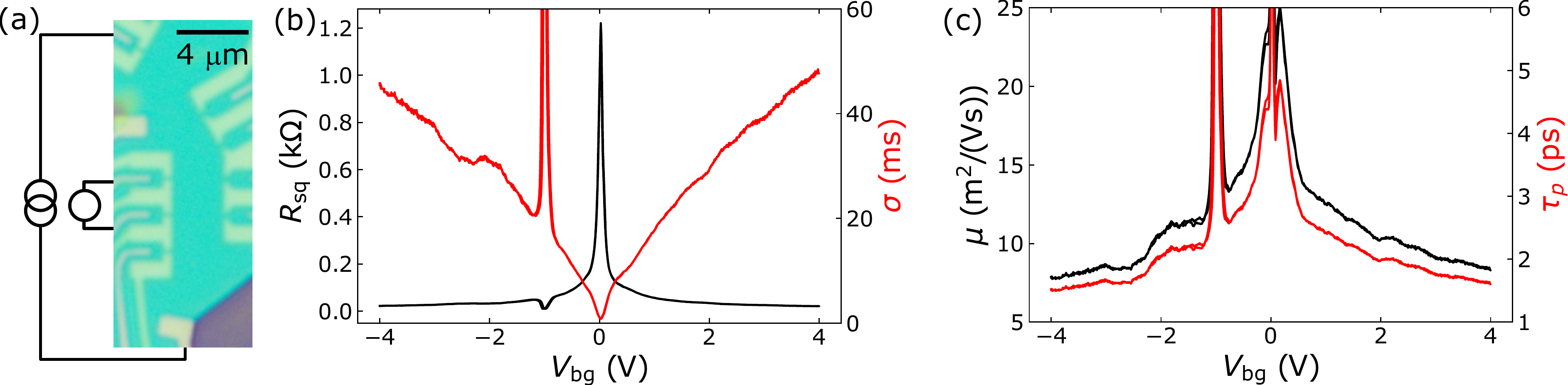}
\caption{(a) Optical microscope image of Sample~1 with the measurement circuit. (b) \Rsq{} and $\sigma=\Rsq{}^{-1}$ vs.~\Vbg{}. (c) Mobility and momentum scattering time vs.~\Vbg{}.\label{FigureRvsVbg}}
\end{figure}

To estimate the device mobility, we have measured the BLG channel resistance as a function of \Vbg{}. The measurement geometry is shown in Fig.~\ref{FigureRvsVbg}a and the result for the square resistance \Rsq{}$=V\times W/(I\times L)$ is shown in Fig.~\ref{FigureRvsVbg}b. Here, $W=7.2\,\mu$m is the BLG width between the $V$ probes and $L=2\,\mu$m is the separation between them. In addition to the expected trend with a peak at the charge neutrality point (CNP) \cite{novoselov2004}, we observe a dip at \Vbg{}$=-1$~V. From the two-point measurements used to characterize the QPCs (which show a peak for \Vbg{}$\approx-1$~V), we conclude that it corresponds to the CNP of the BLG near the contacts. Thus, we attribute the drop in the measured resistance to the gate-tunable invasiveness of the voltage probes, which is minimal for \Vbg{}$\approx-1$~V, when the contact resistance is maximal, resulting in an enhancement of the effective mobility. This explanation is consistent with the channel being ballistic and the contacts having a small overlap with the current path.

We have also plotted $\sigma=\Rsq{}^{-1}$ vs.~\Vbg{}, which does not show the linear trend with \Vbg{} as expected from the Drude model for constant mobility ($\mu$), that predicts $\sigma=ne\mu$. This shows that the effective $\mu$ depends on $n$. To determine $\mu$ (Fig.~\ref{FigureRvsVbg}c) we have used $\mu=\sigma/(ne)$ and used $\mu$ to estimate the momentum scattering time $\tau_p=m^*\mu/e$, where  $m^* = 0.034 \times m_e$ is the effective mass in BLG \cite{mccann2013} and $m_e$ the electron mass. With $v_f\approx0.5\times10^6$~m/s, the mean-free-path $l_\mathrm{mfp}=v_f\times\tau_p\sim 1\,\mu$m, is comparable to the 2~$\mu$m separation between the $V$ probes. This observation indicates that the channel is in the ballistic regime and the measured values represent a lower bound to the actual device quality. The underestimation of $\tau_p$ using this method is confirmed by the clear observation of focusing between QPCs placed at a distance of $L=$4~$\mu$m, requiring a ballistic path of $L\times\pi\approx 12.6$~$\mu$m.
Finally, there is not a significant difference between the electron and hole mobilities for $|\Vbg{}|>1$~V, indicating that the electron-hole asymmetry in the focusing signals is not caused by a difference in mobility.
\FloatBarrier
\section{Two-terminal magnetoresistance at different temperatures}
\begin{figure}[htb]
\centering
\includegraphics[width=\textwidth]{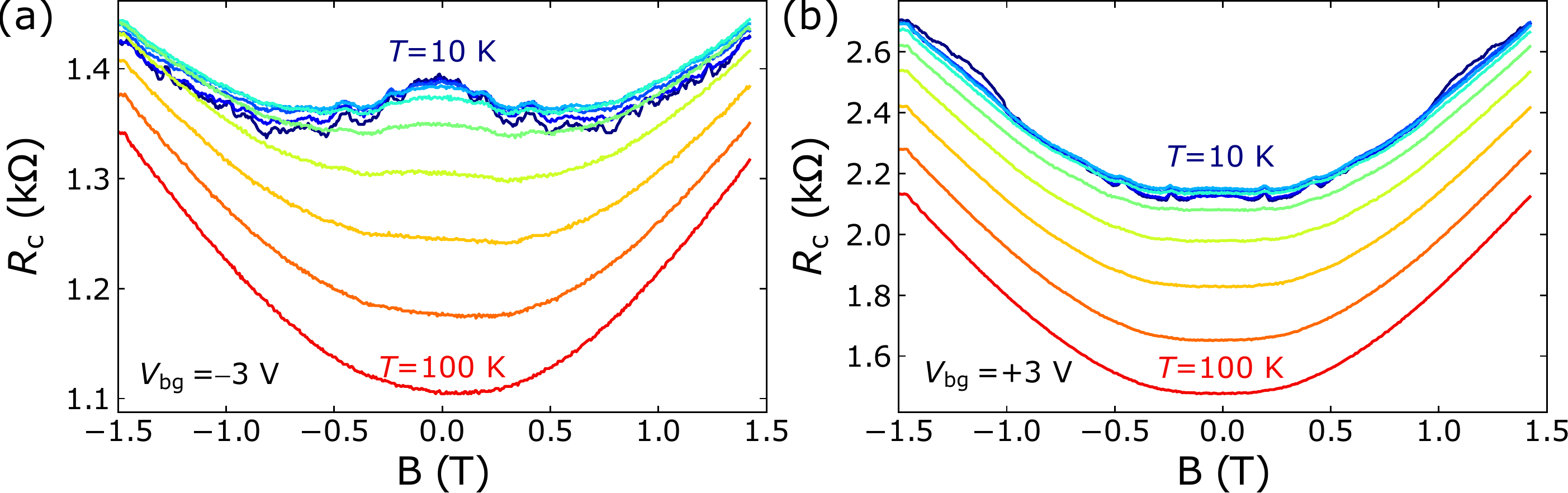}
\caption{Quantum point contact magnetoresistance at \Vbg{}$-3$~V (a) and \Vbg{}$=+3$~V (b).\label{FigureMRvsVbg}}
\end{figure}

To characterize the QPCs, we have measured the $B$-dependence of the two-terminal resistance at \Vbg{}$=\pm3$~V at different $T$. The result is shown in Figs.~\ref{FigureMRvsVbg}a and \ref{FigureMRvsVbg}b and shows small oscillations at low $T$.
There are two main features to highlight from these figures: Firstly, the collector resistance (\Rcol{}) decreases with increasing temperature. This decrease is caused by the small bandgap opening at the BLG under the split gates (around 40~meV for an applied electric field of approximately 0.5~V/nm \cite{icking2022}), that exhibits thermally activated behavior and has to be taken into account to analyze the $T$-dependence of the focusing signal. Secondly, the magnetoresistance below 0.5~T is smaller than 10\% in both cases, allowing us to analyze the low-$B$ focusing peaks assuming that \Rcol{} is constant through the $B$-sweep.

For \Vbg{}$=-3$~V, the low-$T$ data shows a small peak at $B=0$, that is absent for \Vbg{}$=+3$~V. This peak resembles weak localization (WL). Its large width indicates that, if its origin is WL, it must come from a region with a small  phase coherence length (of the 100~nm range). We conclude that it most likely originates from WL near the contacts between the BLG and Ti/Au electrodes, which have been doped by the etching process and are expected to have worse transport properties than the rest of the channel.

\FloatBarrier
\section{Area under peaks}
\subsection{Different temperatures}\label{SectionAreaT}
\begin{figure}[htb]
\centering
\includegraphics[width=0.7\textwidth]{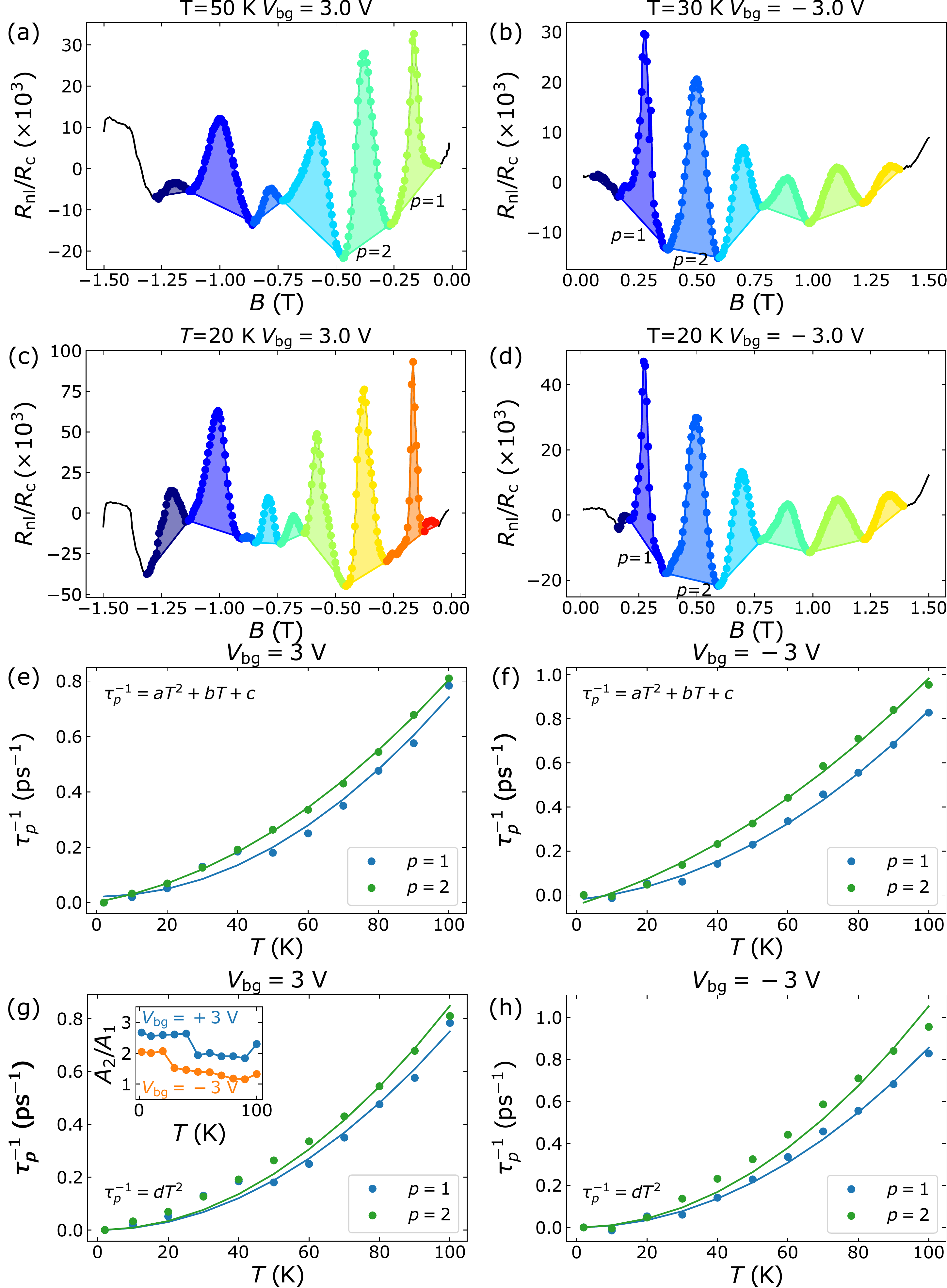}
\caption{(a)-(d) Focusing spectra with the area under the peaks colored. The scatter plot shows the processed data (see text) while the dark line at the background the raw measurement. (e)-(h) Scattering rate $\tau_p^{-1}$ as a function of $T$ from the first two focusing peaks at \Vbg{}$=\pm3$~V. The $p=1$ peak merges with a smaller structure at $T=50 (30)$~K for \Vbg{}$=+3(-3)$~V, leading to a jump in $\tau_p^{-1}$ extracted from this peak. In panels (e) and (f) $\tau_p^{-1}$ vs.~$T$ is fit to $aT^2+bT+c$, whereas in (g) and (h) it is fit to $dT^2$. The fitting parameters can be found in Table~\ref{Table1}. The inset of panel b shows the ratio $A_2/A_1$ vs.~$T$ at \Vbg{}$=\pm3$~V. \label{FigureAreaPeaks}}
\end{figure}
To determine how the scattering rate changes with $T$ we have extracted the area under the focusing peaks shown in Fig.~4 of the main manuscript. This has been done following several steps:
\begin{itemize}
\item \Rnl{} has been normalized by \Rcol{}.
\item A linear background has been corrected from the data.
\item The data has been slightly smoothed (see the small difference between the black curves and colored scatters in Fig.~\ref{FigureAreaPeaks}a-d).
\item The minima in the nonlocal signal have been identified using the find$\_$peaks function from the python package scipy.signal on the reversed data (-$\Rnl{}/\Rcol{}$).

\item The background has been defined for each peak by linear interpolation between the extreme points.

\item The area between the data and background has been calculated using the trapz function from the python package numpy.
\item The scattering rate $\tau_p^{-1}$ has been calculated using \cite{lee2016}:
\begin{equation}
    \tau_p^{-1}=-2v_\mathrm{F}/(\pi L)\log(A_p(T)/A_p(T_\mathrm{base})),
    \label{EqnTp2}
\end{equation}

\end{itemize}
where $v_\mathrm{F}\approx9.0\times10^6$~m/s is the Fermi velocity of BLG at \Vbg{}$=+3$~V, 
$A_p(T)$ is the area under the $p$-focusing peak (starting from $B=0$) at each $T$ and $A_p(T_\mathrm{base})$ is the area of this peak at $T=2$~K.

The result of the process described above is shown in Fig.~\ref{FigureAreaPeaks}a-d at \Vbg{}$=+(-)3$~V and $T=50$, $30$ and $20$~K, respectively. 
The $T$ values have been chosen to illustrate the assimilation of the low-$B$ feature below the $p=1$ peak that affects the $T$ dependence of $\tau_p^{-1}$. Figs. ~\ref{FigureAreaPeaks}e-\ref{FigureAreaPeaks}h show that the $p=1$ and $p=2$ dots overlap for $T\leq40(20)$~K for \Vbg{}$=+(-)3$~V. At higher $T$, the structure shown in Fig~\ref{FigureAreaPeaks}a-\ref{FigureAreaPeaks}d is assimilated by the $p=1$ peak, leading to an increase of the peak area which causes a spurious decrease of $\tau_p^{-1}$. For this reason, we have used the $p=2$ result for our analysis.

In Figs.~\ref{FigureAreaPeaks}e-\ref{FigureAreaPeaks}h we have fit $\tau_p^{-1}$ vs $T$  to two parabolas: $\tau_p^{-1}=aT^2+bT+c$ (Figs.~\ref{FigureAreaPeaks}e and \ref{FigureAreaPeaks}f), and $\tau_p^{-1}=aT^2$ (Figs.~\ref{FigureAreaPeaks}g and \ref{FigureAreaPeaks}h). The former fits the $p=2$ result better for both $\Vbg=+3$~V and $-3$~V. 

A quadratic $T$-dependence of $\tau_p^{-1}$ is associated with electron-electron interactions \cite{lee2016}. In contrast, a linear dependence is associated with phonon-dominated scattering \cite{taychatanapat2013,hwang2008}. Thus, our analysis indicates that both scattering terms are relevant. By calculating $T_0=b/a$, which is the $T$ where the quadratic term starts to dominate over the linear term, we obtain $T_0=37\pm3$ $(90\pm20)$~K for \Vbg{}$=+(-)3$~V.  
\FloatBarrier

Since we are analyzing the $p=2$ peaks, it would be tempting to attribute the electron-hole asymmetry to diffuse scattering at the edge (DSE), that the TEF spectra indicates may be stronger for holes. If DSE was $T$-dependent, it could lead to a faster $p=2$ peak decay with $T$ and artificially enhance $t_p^{-1}$. By monitoring the $T$-dependence of $A_2/A_1$ we can determine whether the $bT$ term is dominated by $T$-dependent DSE because, in this case, $A_2/A_1$ would decrease linearly with increasing $T$. We have plotted $A_2/A_1$ at the inset of Fig.~\ref{FigureAreaPeaks}b and found that the most clear feature is a sudden drop at $T=50$ $(30)$~K for \Vbg{}$=+3$ $(-3)$~V, as expected from Fig.~\ref{FigureAreaPeaks}, indicating that the dominant scattering source giving rise to the TEF amplitude decay with $T$ at \Vbg{}$=+3$~V and the linear $T$-dependence of $\tau_p^{-1}$ is not DSE. Note that the 100~K case shows a clear difference with respect to the others. We attribute it to the thermally-activated transport across the weakly-gaped ($\sim40$~meV) BLG region at 100~K where $k_BT\approx8.6$~meV.
\begin{table}
\caption{Fitting parameters obtained from the temperature dependence of $\tau_p$ using $\tau_p^{-1}=aT^2+bT+c$ (Figs.~\ref{FigureAreaPeaks}e and \ref{FigureAreaPeaks}f) and $\tau_p^{-1}=dT^2$ (Figs.~\ref{FigureAreaPeaks}g and \ref{FigureAreaPeaks}h).\label{Table1}}
\begin{ruledtabular}
\begin{tabular}{ c c c c c c }
 $ $& $ $ & $a$ (ps$^{-1}$K$^{-2}$) & $b$ (ps$^{-1}$K$^{-1}$) & $c$ (ps$^{-1}$) & $d$ (ps$^{-1}$K$^{-2}$)\\
\hline
\Vbg{}$=+3$~V & $p=1$ & $(7\pm 1)\times10^{-5}$  & $(-0.09\pm 1)\times10^{-3}$& $(2\pm 3)\times10^{-2}$&$(7.5\pm0.2)\times10^{-5}$\\
\hline
              & $p=2$ & $(5.9\pm 0.3)\times10^{-5}$& $(2.2\pm 0.3)\times10^{-3}$& $(3\pm 6)\times10^{-3}$&$(8.5\pm0.2)\times10^{-5}$ \\
\hline
\Vbg{}$=-3$~V & $p=1$ & $(7.1\pm 0.6)\times10^{-5}$  & $(1.5\pm 0.7)\times10^{-3}$& $(-4\pm 2)\times10^{-2}$&$(8.6\pm0.1)\times10^{-5}$\\
\hline
              & $p=2$ & $(5.5\pm 0.8)\times10^{-5}$& $(4.8\pm 0.9)\times10^{-3}$& $(-4\pm 2)\times10^{-2}$&$(1.05\pm0.03)\times10^{-4}$ \\
\end{tabular}
\end{ruledtabular}
\end{table}
\subsection{Base temperature}
\begin{figure}
\centering
\includegraphics[width=\textwidth]{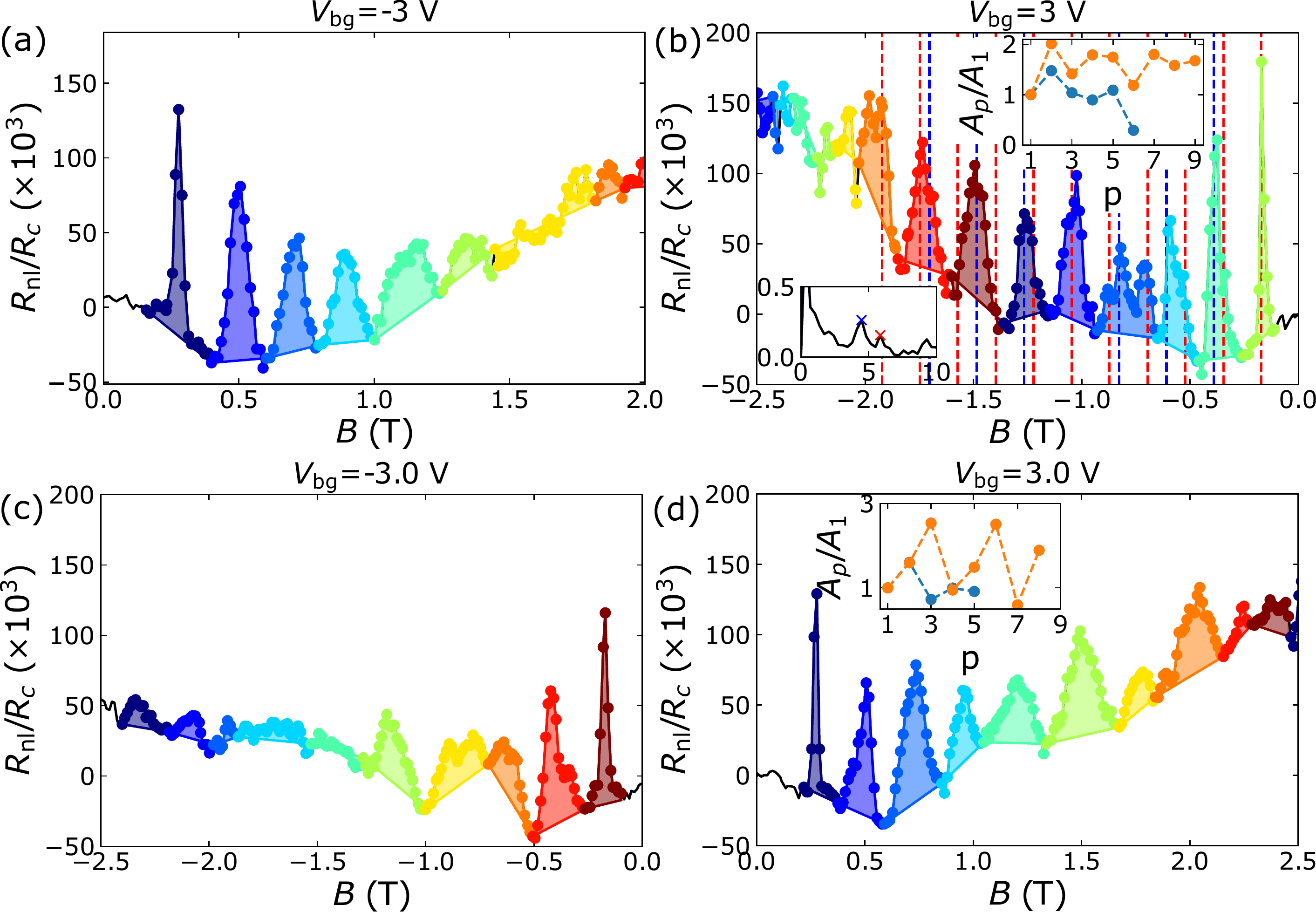}
\caption{Focusing spectra obtained in configuration C1 and in C2 at \Vbg{}$=+3$~V (a), (c) and \Vbg{}$=-3$~V (b), (d), respectively. The area under the peaks is colored and the scatter plot shows the processed data while the dark line at the background the raw measurement (see text). The upper insets in panels b and d show the area under peak p ($A_p$) normalized by the area under the first peak ($A_1$) for \Vbg{}$=-3$~V (orange) and \Vbg{}$=+3$~V (blue). The equally-spaced blue and red vertical lines in panel b indicate the expected peak positions according to the Fourier analysis, which is shown at the lower inset of panel b. The $y$-axis is the amplitude of the Fourier transform of the TEF spectrum and the $x$-axis the Fourier frequency. The selected peaks are marked by crosses and color-coded according to the vertical lines. \label{FigureAreaPeaks2K}}
\end{figure}
To determine the specularity of the edge reflection in the TEF measurements, we have determined the area under the different peaks in the \Rnl{} data shown in Fig.~2c of the main manuscript. To take into account the contact magnetoresistance shown in Fig.~\ref{FigureMRvsVbg} we have normalized \Rnl{} by $R_c$ using the data from Fig.~\ref{FigureMRvsVbg} at $T=10$~K.
The result, obtained following the procedure described in Section~\ref{SectionAreaT}, is shown in Fig.~\ref{FigureAreaPeaks2K} and shows that, for \Vbg{}$=-3$~V, the peak amplitude decays much faster than for \Vbg{}$=+3$~V, as shown in the main manuscript. Additionally, the amplitude of the \Vbg{}$=+3$~V signal is still around 25\% larger than the \Vbg{}$=-3$~V case. We attribute this small difference to the fact that the detector has a slightly larger asymmetry at \Vbg{}$=\pm3$~V than the injector, which is the contact we corrected for. 
At the inset of Fig.~\ref{FigureAreaPeaks2K}b we show the normalized area under the different peaks in both \Vbg{}$=-3$~V and \Vbg{}$=+3$~V cases where one can see more clearly the faster decrease of peak amplitude in the former.

An additional feature which can be identified in the \Vbg{}$=+3$~V data is the apparent beating pattern which results in the splitting of the $p=4$ TEF peak. To infer whether it is compatible with the expected interference pattern arising from TEF between QPCs which are slightly misaligned with respect to a crystallographic direction, we calculated the Fourier transform of the signal. The result, shown at the lower inset of Fig.~\ref{FigureAreaPeaks2K}b, indicates that two clear peaks are present at the expected FFT frequency range. Because the measured $B$ fields are not exactly equally spaced, the measured data has been mapped on a $B$ axis with equally spaced points by interpolation from the raw TEF data. Because the TEF peaks are narrow, details on the mapping such as small offsets in $B$ can modify the FFT peak shapes. To correct for this issue we used a mesh of 1k equally spaced points from 0.1 to 3~T. As a result the FFT peak positions are robust against $B$ shifts up to 0.1~T. 
The peaks obtained at the expected frequencies are marked by a red and a blue cross and their frequencies are inverted to determine the corresponding periodicities, which are represented as vertical dashed lines. The result, which is compatible with the first six peaks, indicates that a beating pattern with two separate frequencies can explain most of the spectra obtained at \Vbg{}$=+3$~V. 

\FloatBarrier

\section{Transverse electron focusing at different geometries}\label{SectionTEFGeoms}
\begin{figure}
\centering
\includegraphics[angle=0,width=\textwidth]{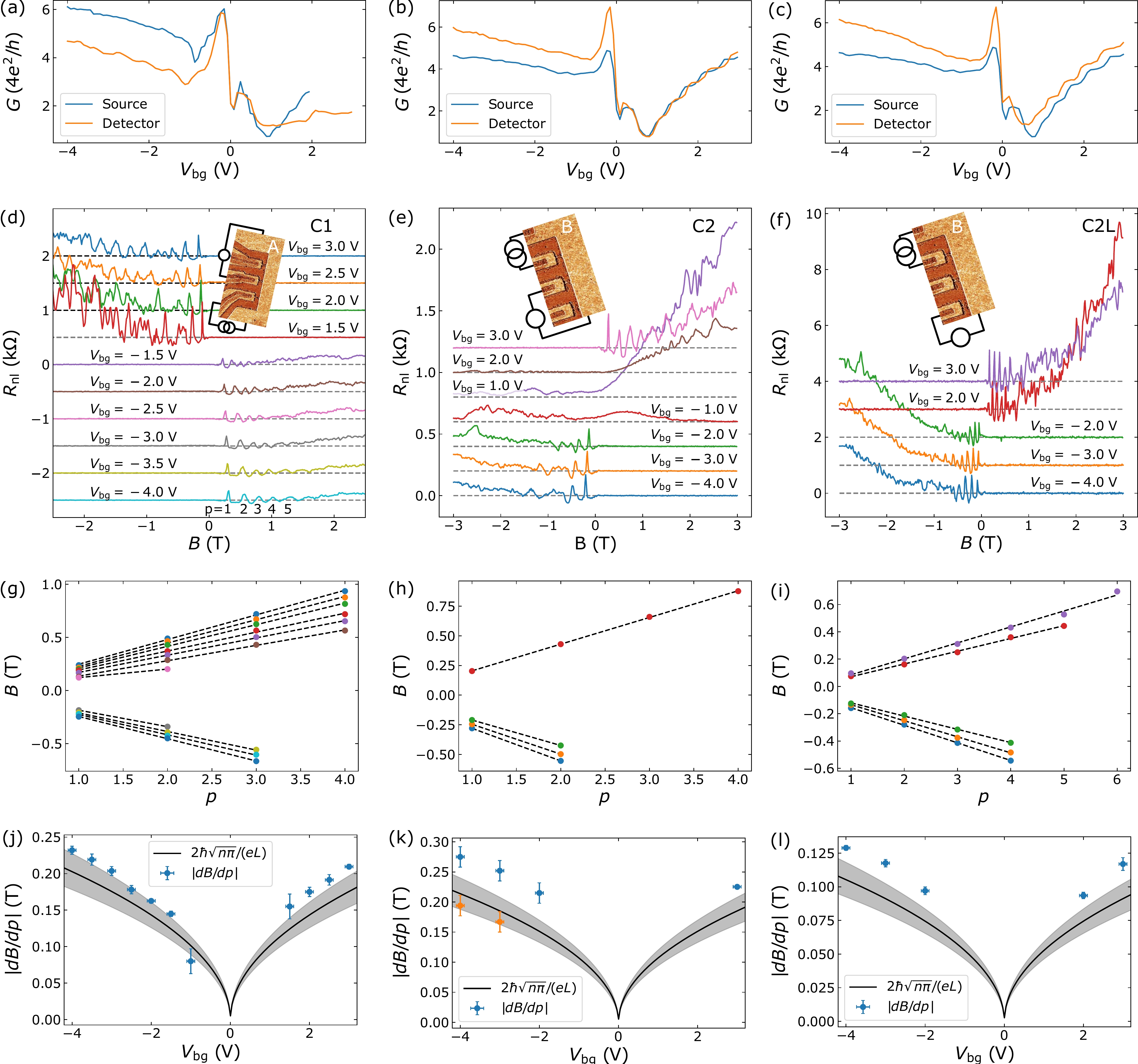}
\caption{TEF measurements obtained using the QPCs in rectangles A and B of Fig.~\ref{FigureAFM}c. (a-c) $G$ vs.~\Vbg{} calculated as in the main manuscript for the QPCs used in the TEF measurements shown in (d-f), respectively. The insets correspond to the measurement geometries represented with the same orientation as in Fig.~\ref{FigureAFM}c for clarity. (g-i) Peak positions vs.~$p$ at different \Vbg{}. The dashed lines are the fits to $B=B_0+(dB/dp)\times p$, where $B_0$ accounts for the magnet remanence and $dB/dp$ is the slope. (j-l) Slope of fits obtained from panels g-i, respectively, together with a fit to Equation~\ref{EquationTEF} assuming normal incidence ($\theta=0$; dashed lines). \label{FigureClosedQPCs}}
\end{figure}
\begin{figure}
\centering
\includegraphics[angle=0,width=0.66\textwidth]{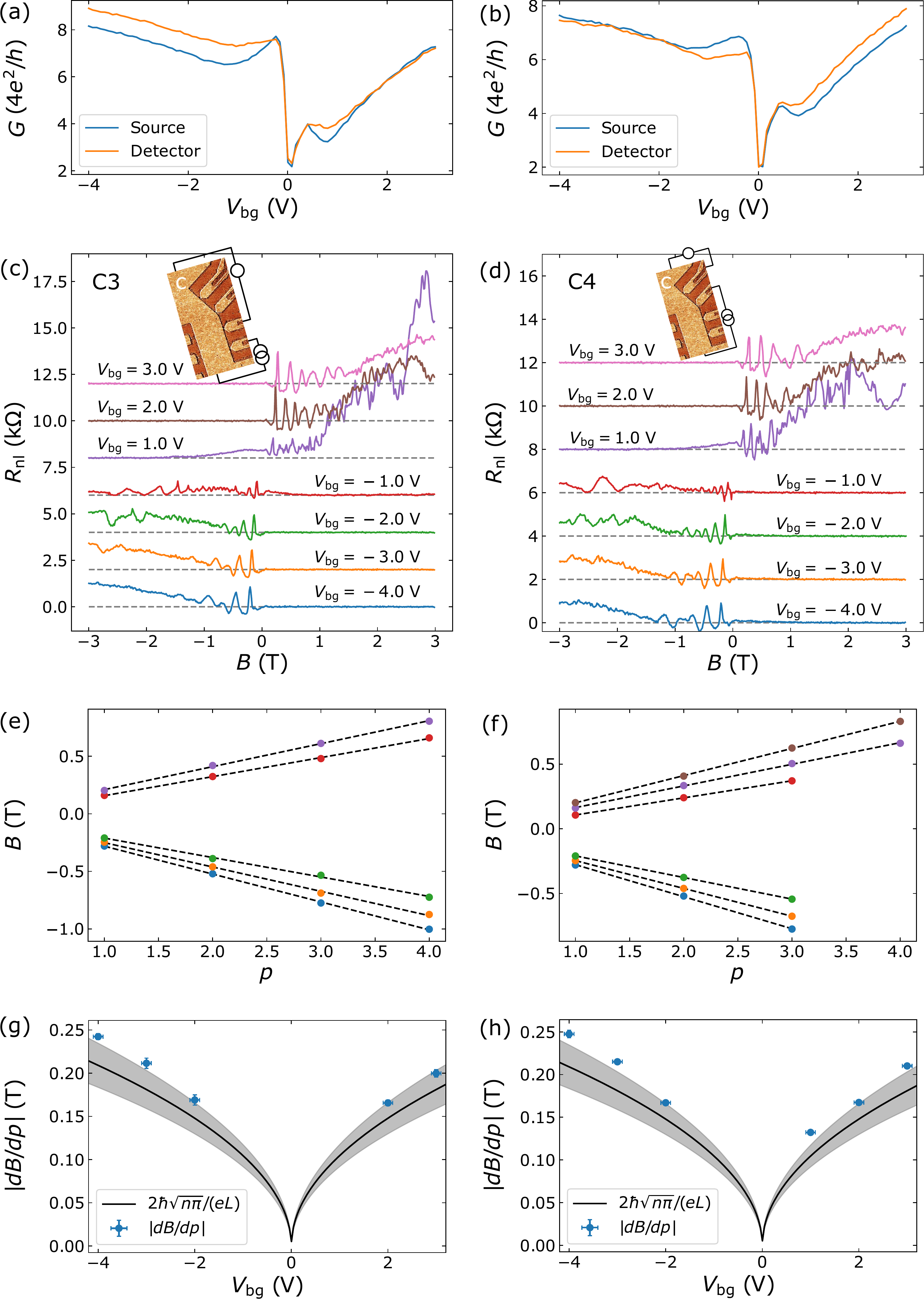}
\caption{TEF measurements obtained using the QPCs in rectangle C of Fig.~\ref{FigureAFM}c. (a) and (b) $G$ vs.~\Vbg{} calculated as in the main manuscript for the QPCs used in the TEF measurements shown in (c) and (d), respectively. The insets correspond to the measurement geometries represented with the same orientation as in Fig.~\ref{FigureAFM}c for clarity. (e) and (f) Peak positions vs.~$p$ at different \Vbg{}. The dashed lines are fits to to $B=B_0+(dB/dp)\times p$. (g) and (h) Slope of the  fits obtained from panels e and f, respectively, together with a fit to Equation~\ref{EquationTEF} assuming normal incidence ($\theta=0$; dashed lines). \label{FigureOpenQPCs}}
\end{figure}
In the presence of trigonal warping, the transverse electron focusing (TEF) spectra are expected to depend on the relative orientation of the QPCs with respect to the crystallographic directions of the BLG \cite{gold2021, manesco2022}. 
For this reason, we have studied TEF using gate-defined QPCs which are oriented in different directions. In particular, the geometry used to obtain the TEF spectra shown in the main manuscript, which we also show in Fig.~\ref{FigureClosedQPCs}d, has a rotation of 30$^\circ$ with respect to the second set of QPCs, which is shown in Fig.~\ref{FigureClosedQPCs}e (see Fig.~\ref{FigureAFM}c for an overview of the whole sample). Since the QPCs in area B (Fig.~\ref{FigureAFM}c) are oriented along a straight edge in the BLG, which is most likely a crystallographic direction \cite{neubeck2010}, the QPCs in area A are expected to be aligned along an armchair (or zig-zag) direction while the ones in area B must be along a zig-zag (armchair) direction, although we cannot tell which one is which.

Comparing Fig.~\ref{FigureClosedQPCs}d with  Fig.~\ref{FigureClosedQPCs}e, the spectra occur at opposite $B$. This happens because in Fig.~\ref{FigureClosedQPCs}d the current source is at the left of the voltage probe whereas in Fig.~\ref{FigureClosedQPCs}e it is at the right. For negative \Vbg{} in Fig.~\ref{FigureClosedQPCs}e there are fewer peaks than in Fig.~\ref{FigureClosedQPCs}d and the $p=2$ peak is split into two. 

To find if the focusing peaks occur at the expected $B$, we use  \cite{vanHouten1989}:
\begin{equation}
\Bfocus{}=\frac{2p\hbar k_F\cos{\theta}}{eL}=\frac{2p\hbar \sqrt{n\pi}\cos{\theta}}{eL},
\label{EquationTEF}
\end{equation}
where $\hbar$ is the reduced Plank constant, $k_f$ the Fermi wavevector, $\theta$ the electron incidence angle from the QPC with respect to the normal, and $L$ the contact separation. Note that Equation~\ref{EquationTEF} corresponds to Equation~1 of the main manuscript. 

To compare Equation~\ref{EquationTEF} with the measured data, we have identified \Bfocus{} as the $B$ values where \Rnl{} is maximal, plotted \Bfocus{} vs.~$p$ at each \Vbg{}, and fitted the data to $\Bfocus{}=B_0+(dB/dp)\times p$, where $B_0$ accounts for the coercivity of the magnet and $dB/dp$ quantifies the change in \Bfocus{} with $p$ (see Figs.~\ref{FigureClosedQPCs}g-i).
After this, we have plotted $|dB/dp|$ vs.~$n$ and compared this result with Equation~\ref{EquationTEF} assuming $\cos(\theta)=1$, which provides the maximal peak separation (see Figs.~\ref{FigureClosedQPCs}j-l). We observe that the separation between the measured peaks is even larger than predicted by Equation~\ref{EquationTEF}.
 This result is more pronounced in Fig.~\ref{FigureClosedQPCs}k, and may be consistent with an impurity obstructing the charge transport path and leaving a smaller peak at $B<\Bfocus{}$ (which we did not consider in Fig.~\ref{FigureClosedQPCs}h) and a larger peak at $B>\Bfocus{}$. As a result the spacing between the first and second peaks is artificially enhanced and, since the $p=3$ peak is not clear enough, $|dB/dp|$ is overestimated. The opposite occurs when we consider the smaller peaks at lower $B$ (Fig.~\ref{FigureClosedQPCs}k, orange dots). 
 
For \Vbg{}$>0$ the background signals in Fig.~\ref{FigureClosedQPCs}e are much larger than in  Fig.~\ref{FigureClosedQPCs}d and additional features that are not expected from TEF are observed. We attribute them to quantum interference and restrict our comparison to \Vbg{}$=3$~V. In this case, up to eight TEF peaks are visible, as in  Fig.~\ref{FigureClosedQPCs}d, but without the splitting of the $p=4$ peak. Looking at the spacing between the peaks, we see that in both cases it is close to the fit for $\cos(\theta)=1$, indicating that there is only one current jet which departs at normal incidence from the QPCs.

For completeness, we have also measured TEF over a distance of 4~$\mu$m using the same injector as in  Fig.~\ref{FigureClosedQPCs}e but connecting the detector to the lowest QPC electrode. The result is shown in  Fig.~\ref{FigureClosedQPCs}f and, as expected from Equation~\ref{EquationTEF}, the peak spacing is approximately halved with respect to  Fig.~\ref{FigureClosedQPCs}e. The result is summarized by comparing  Fig.~\ref{FigureClosedQPCs}k with  Fig.~\ref{FigureClosedQPCs}l. We note, however, that in both cases the spacing obtained from the spectra is slightly larger than predicted by the model.

To study the influence of the size quantization of the QPCs on the TEF data we have also prepared some QPCs with a horn-like shape which do not show size quantization for any \Vbg{} (see Figs.~\ref{FigureOpenQPCs}a and \ref{FigureOpenQPCs}b). The TEF spectra shown in Figs.~\ref{FigureOpenQPCs}c and \ref{FigureOpenQPCs}d looks similar to the data shown in \ref{FigureClosedQPCs} with similar spacing between peaks (Figs.~\ref{FigureOpenQPCs}g and \ref{FigureOpenQPCs}h), and 6 peaks visible in the best case (Fig.~\ref{FigureOpenQPCs}c, \Vbg{}$=+3$~V). Near the CNP, quantum interference features, similar to those observed in Figs.~\ref{FigureOpenQPCs}a and \ref{FigureClosedQPCs}, are observed indicating that size quantization at the QPCs does not play an important role in the TEF spectra. 
As shown in Figs.~\ref{FigureOpenQPCs}c and \ref{FigureOpenQPCs}d, inset, the electrodes used for TEF in the C3 and C4 geometris are also rotated by 30$^\circ$. We thus believe that the close similarity between the TEF spectra obtained in both cases indicates that trigonal warping does not play a dominant role in our measurements.
\FloatBarrier
\subsection{Reciprocity}
\begin{figure}
\centering
\includegraphics[width=\textwidth]{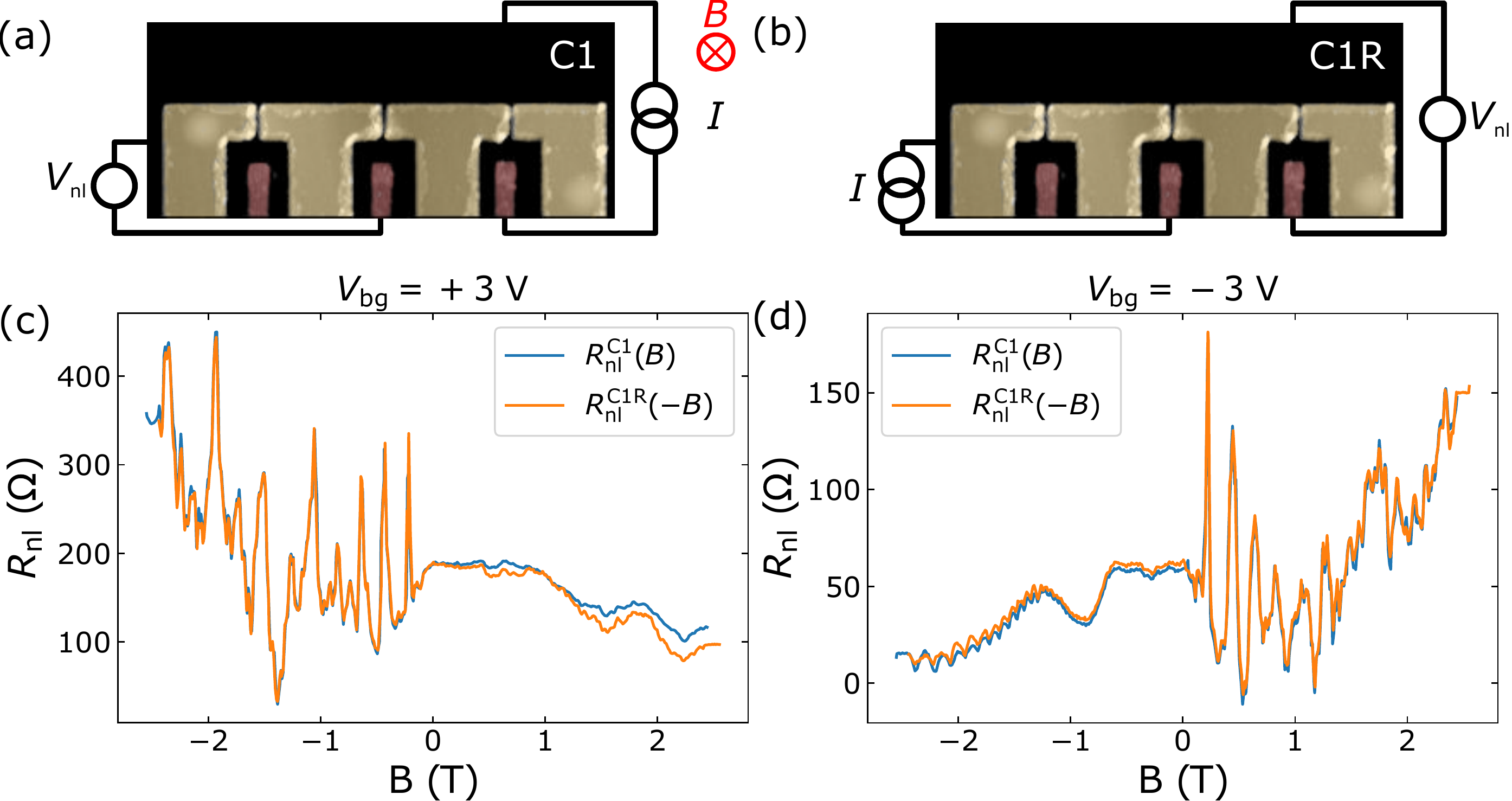}
\caption{(a) and (b) Measurement geometries C1 and C1R used to measure the reciprocity of the TEF data, which is shown in panels (c) and (d) for \Vbg{}$=+3$~V and $-3$~V, respectively. The $B$ has been reversed for the C1R measurements to show the agreement between both curves in a more clear way.\label{FigureReciprocity}}
\end{figure}

To confirm that the TEF measurements are in the linear response regime we have measured \Rnl{} in C1 ($\Rnl^\mathrm{C1}$) and its reciprocal geometry (C1R, $\Rnl^\mathrm{C1R}$), obtained by swapping the current and voltage leads \cite{buttiker1986} (Figs.~\ref{FigureReciprocity}a and \ref{FigureReciprocity}b). 
 The measured data is shown in Figs.~\ref{FigureReciprocity}c and \ref{FigureReciprocity}d for \Vbg{}$=+3$~V and $-3$~V, respectively. To facilitate the comparison between the signals $\Rnl^\mathrm{C1}$ and $\Rnl^\mathrm{C1R}$ we have represented $\Rnl^\mathrm{C1R}$ as a function of $-B$. The almost perfect overlap of the two focusing spectra in both \Vbg{} confirms that our measurements are in the linear response regime. We attribute the small differences to electrostatic changes originated from sweeping the gate voltages.
 
\begin{figure}
\centering
\includegraphics[width=\textwidth]{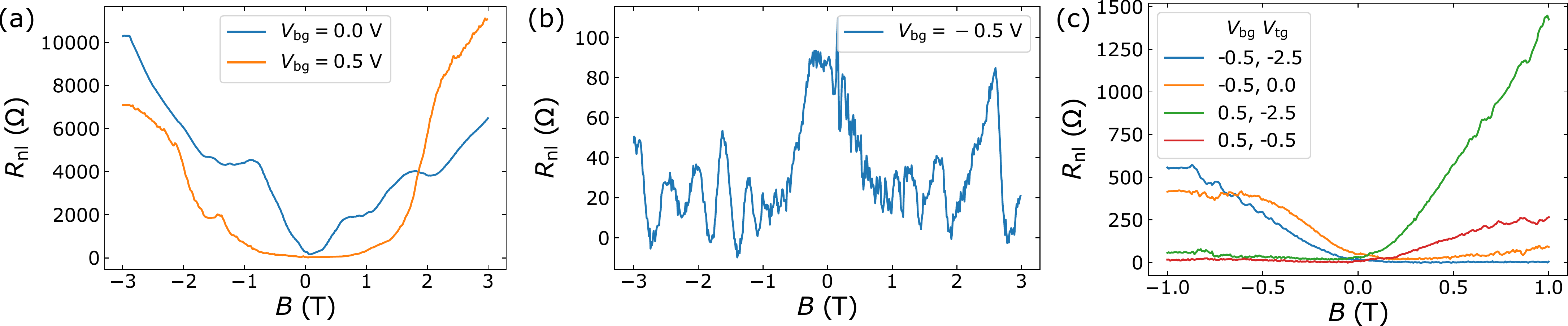}
\caption{\Rnl{} near the charge neutrality point using geometry C1. (a) and (b) were obtained setting the top-gated regions to $n=0$ and (c) was obtained with both regions doped, inducing p-n junctions in the BLG channel (see text). \label{FigureFocusingCNP}}
\end{figure}
\FloatBarrier
\section{Transverse electron focusing near the charge neutrality point}

When measuring \Rnl{} at $|\Vbg{}|<1$~V the double-gated regions have a very small band gap and play a role in charge transport. We have measured \Rnl{} vs.~$B$ in configuration C1 of the main manuscript and the results are shown in Fig.~\ref{FigureFocusingCNP}. In  Figs.~\ref{FigureFocusingCNP}a and \ref{FigureFocusingCNP}b, \Vtg{} has been set to keep the double-gated regions at $n\approx0$. In Fig.~\ref{FigureFocusingCNP}c we show \Rnl{} in the presence of p-n junctions (orange and green lines) with both the single-gated and double-gated regions being n (red) and p doped (blue). 
The main result from Fig.~\ref{FigureFocusingCNP}c is the presence of plateaux which are clearly visible for $|B|$ as small as 0.5~T.
 This observation may be explained by considering that, when the top-gated regions conduct, the nonlocal geometry behaves like a Hall geometry when $B$ deflects the injected carriers towards the detector. In this condition, if the system enters the quantum hall regime, one would expect plateaux. However, a significant amount of the observed plateaux occur at fields where the Shubnikov-de Haas oscillations are not yet well developed (Fig.~\ref{SdHOscillations}), indicating that additional effects may be at play, such as quantum interference. In this case, oscillations superimposed on the linear Hall-like signal may look like plateaux on \Rnl{}. The role of quantum interference is specially clear for the result in Fig.~\ref{FigureFocusingCNP}b, where the linear background is not observed and \Rnl{} oscillates both for positive and negative $B$. We believe that the symmetric $B$-dependence, which is neither consistent with the BLG being electron or hole doped, can be explained considering that the top-gated regions are charge neutral and very close to the zero electric field condition, implying no bandgap opening and that they can conduct current (see color map in Fig.~1b of the main manuscript). A similar behavior is observed in Fig.~\ref{FigureFocusingCNP}a. In this case, an additional effect occurs: \Rnl{} increases dramatically when $|B|>1.5$~T. We believe that this is due to the formation of an insulating state in BLG near the CNP at high $B$ \cite{feldman2009}.

\FloatBarrier

\section{Transverse electron focusing Sample~2}
\begin{figure}
\centering
\includegraphics[width=\textwidth]{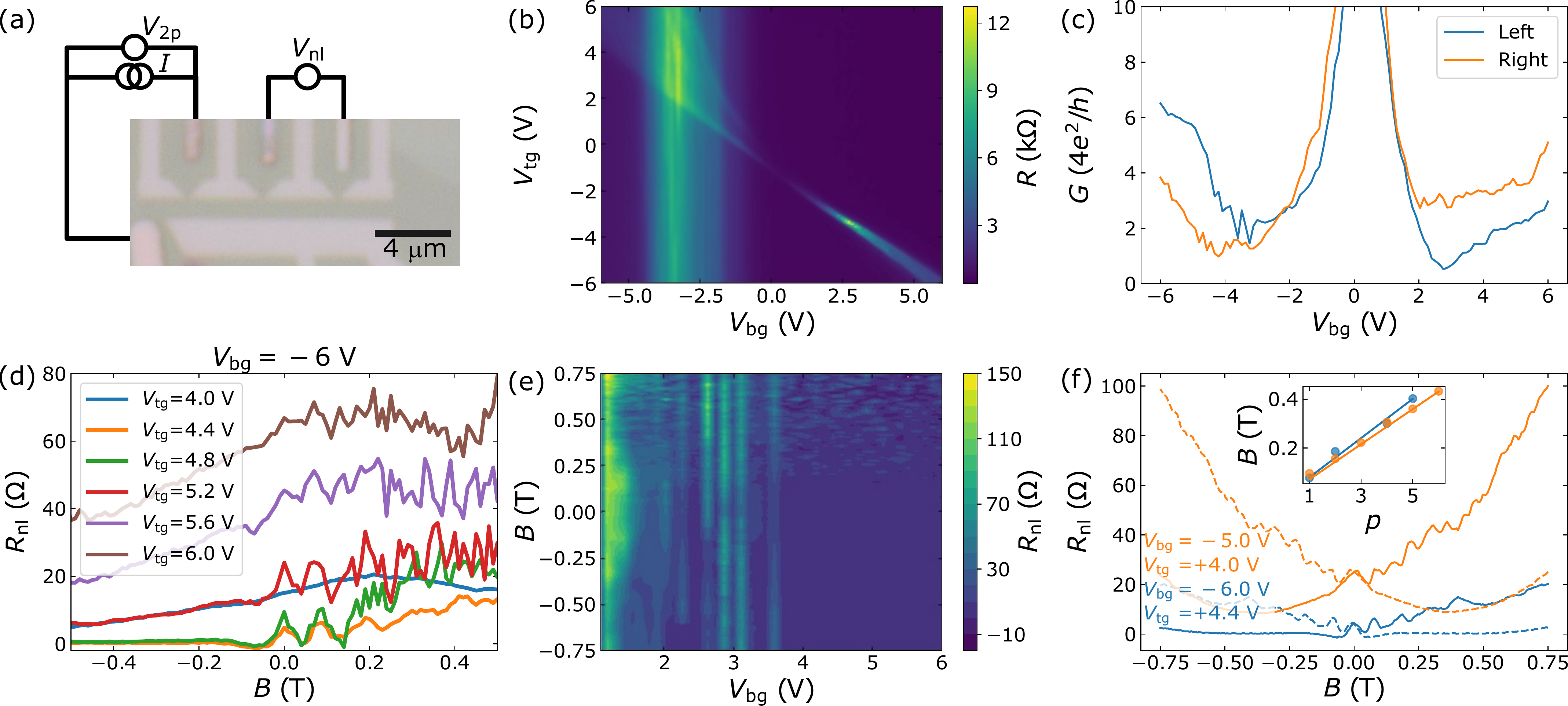}
\caption{(a) Optical microscope image of Sample~2. The contacts to the BLG are connected to the $V$ probes and $I$ sources and the top gates have a horn-like shape. The measurement circuits are represented using black lines. (b) Two-point resistance $R=V_\mathrm{2p}/I$ as a function of \Vbg{} and \Vtg{}. (c) Conductance of the left and middle QPCs vs.~\Vbg{} extracted as in Fig.~1 of the main manuscript. (d) $\Rnl{}=\Vnl{}/I$ vs.~$B$ at \Vbg{}$=-6$~V and at different \Vtg{}, showing clear features of TEF. (e) \Rnl{} vs.~$B$ for positive \Vbg{} and the \Vtg{} that brings the double gated regions to the CNP. (f) \Rnl{} vs.~$B$ at \Vbg{}$=-6$~V and $-5$~V. The dashed lines show the data obtained in the reciprocal geometry where we replaced the $I$ source with the $V$ measurement module. The inset shows the experimental (dots) and theoretical (lines) \Bfocus{}. The latter are obtained from Equation~\ref{EquationTEF} with $\theta=0$ and $n$ obtained from Shubnikov de Haas oscillations. The data from panels (b)-(e) was obtained at 640~mK and panel (f) at 2.4~K. \label{FigureSample2}}
\end{figure}
We have also measured TEF in a second heterostructure. Its top and bottom hBN thicknesses are 49 and 40~nm, respectively, and its optical microscope image is shown in Fig.~\ref{FigureSample2}a. The QPC resistance ($R=V_\mathrm{2p}/I$, where $V_\mathrm{2p}$ is the two-terminal voltage and $I$ the applied current, defined in Fig.~\ref{FigureSample2}a) is shown in Fig.~\ref{FigureSample2}b as a function of \Vbg{}, applied to a multilayer graphene backgate and \Vtg{}, applied to the top gates surrounding the left QPC. Following the protocol described in the main manuscript, we extracted $G$ of the left and middle QPCs in Fig.~\ref{FigureSample2}a using $G = (R_\mathrm{max}-R_\mathrm{min})^{-1}$, where $R_\mathrm{max(min)}$ is the maximum (minimum) value of $R$ at each \Vbg{}. The result is shown in Fig.~\ref{FigureSample2}c. Note that, since the hBN thicknesses are approximately two times thicker than for Sample~1, the applied \Vbg{} and \Vtg{} are also larger. 

By monitoring $\Rnl{}=\Vnl{}/I$ vs.~$B$, where \Vnl{} is the nonlocal voltage, we obtained the TEF spectra shown in Fig.~\ref{FigureSample2}d for \Vbg{}$=-6$~V and at different \Vtg{}. This result shows that, even though the background is sensitive to small changes on \Vtg{}, the first TEF peak is robust against a \Vbg{}$=0.2$~V change. Note that \Rnl{} shows a peak for $B=0$ which is not expected from TEF measurements. The results obtained for \Vbg{}$>0$ are shown in Fig.~\ref{FigureSample2}e and, even though signatures of quantum interference can be seen for $B>0.25$~T, no clear signs of TEF can be found. 

To confirm that our measurements are in the linear response regime we performed reciprocity checks on the TEF data measured at \Vbg{}$=-5$~V and $-6$~V. The almost perfect match between both curves confirms that the bias current (100~nA) is small enough so that our measurements are in the linear response regime.  

The TEF spectra shown in Figs.~\ref{FigureSample2}d and \ref{FigureSample2}f show a peak at $B\approx0$ which would be compatible with the $p=1$ TEF peak if the magnet remanence $B_0\approx-$\Bfocus{}. This hypothesis can be ruled out from the reciprocity data. A large $B_0$ would lead to a horizontal ($B$) shift of the reciprocal measurements (dashed lines) with respect to the original ones (solid lines). This is a consequence of the reciprocity theorem which states that, in a non-magnetic system and in the linear response regime, $R_{ijkl}(B)=R_{klij}(-B)$, where the first pair of indexes denote the contacts connected to the $I$ source and the second pair the contacts used for the $V$ measurements \cite{buttiker1986}. If $B$ is shifted by $B_0$, then $R_{ijkl}$ will coincide with $R_{klij}$ at $B=B_0$ instead of 0. Note that both measurements were performed sweeping the magnet from $750$~mT to $-750$~mT so the same $B_0$ is expected from both measurements. Thus, the coincidence of the direct and reciprocal TEF data at $B\approx0$ confirms that $B_0\approx 0$ and we can conclude that \Rnl{} shows an additional peak at $B=0$. Most likely, this peak is caused by the detection of the ballistic electron stream reflected at the opposite BLG edge.    
\FloatBarrier
\section{Numerical simulations}

We performed semiclassical calculations of electron focusing in BLG.
These calculations require previous knowledge of the angular distribution of the currents and the shape of the Fermi surface. These parameters were obtained using a tight-binding model implemented in Kwant~\cite{groth2014}.
Both calculations are explained below and the code is available at \cite{zenodo}.

\subsection{Tight-binding model}

\begin{figure}
\centering
\includegraphics[width=\textwidth]{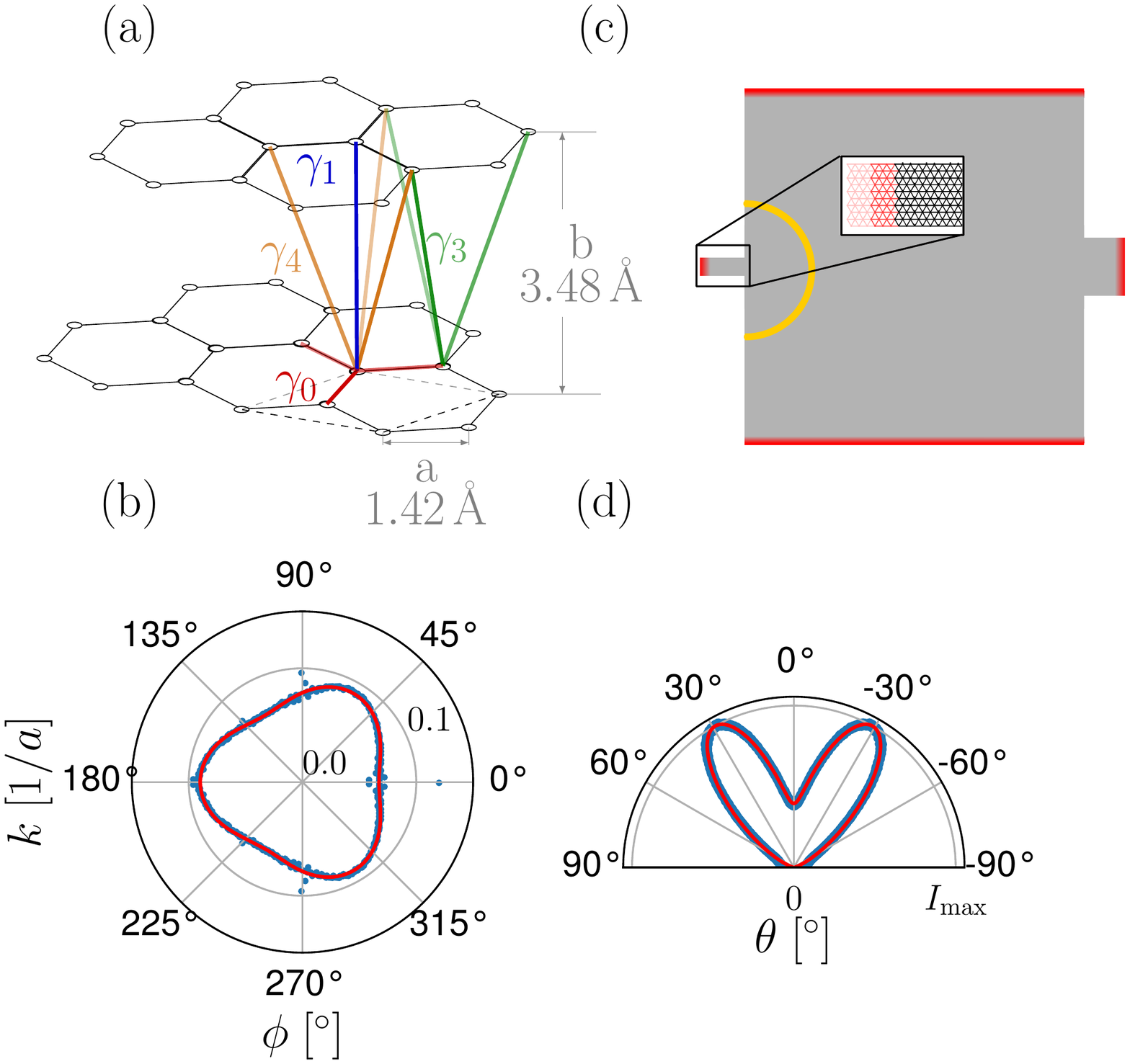}
\caption{
    (a) Illustration of the BLG tight-binding model with the relevant hoping parameters $\gamma_{0}$ to $\gamma_{4}$. The unit cell is defined by the dashed lines at the lower graphene layer.
    (b) Computed Fermi surface (blue dots) and fit from Eq.~\ref{FermiSurfaceGr} (red line).
    (c) Device simulated to extract the current angle distribution.
    Electrons are injected from the QPC in the left, and the current density is computed at the orange line.
    The extra leads avoid backreflection of the electrons to the orange circle.
    The leads are shown in red.
    (d) Computed current distribution (blue dots) and fitted distribution from Eq.~\ref{EquationDist} (red line).
    \label{FigNumerics}
}
\end{figure}

We implement the tight-binding model from \cite{jung2014,mccann2013} which includes four hopping parameters:
$$
\mathcal{H} = -\mu\sum_{n, l} c_{n, l}^{\dagger}c_{n, l} + \sum_{i=0}^4\sum_l \sum_{n, m \in S_i} \gamma_i c_{m, l}^{\dagger}c_{n, l} + \Delta \sum_n (c_{n,1}^{\dagger}c_{n,1} - c_{n,2}^{\dagger}c_{n,2})~,
$$
where $c_{n, l}$ and $c_{n, l}^{\dagger}$ are the annihilation and creation operators for electron states at position $n$ and layer $l$, $\mu$ is the chemical potential, $\Delta$ is the layer imbalance, and the sets of hoppings $S_i$, with corresponding strength $\gamma_i$, are shown in Fig.~\ref{FigNumerics}a.
Note that $\gamma_2=0$ in BLG~\cite{jung2014} and is not shown in the figure.

From this model, we extract the Fermi surface used in the semiclassical calculations.
To reproduce the experimental conditions, we use the displacement field and electron density corresponding to the curves with \Vbg{=+3V} in Fig.~1.
From the experimental data, we find the corresponding tight-binding parameters $\mu=98$~meV, and $\Delta=84$~meV. 
We fit the corresponding Fermi surface with the lowest Fourier component that accounts for trigonal warping. Namely,
\begin{equation}
k_{\tau}(E, \phi)=k_{F,0}+\tau \delta k \sin(3\phi + \phi_c), 
\label{FermiSurfaceGr}
\end{equation}
where $\tau=\pm1$ in valley $\pm$K, $\phi$ is the polar angle, and $k_{F, 0}$, $\delta k$, and $\phi_c$ are the fitting parameters.
We show the computed and fitted data in Fig.~\ref{FigNumerics}b.

Finally, because of trigonal warping, it is incorrect to assume that the injected electrons have an uniform angular distribution.
Instead, electrons are injected as two valley-polarized jetstreams from the QPC.
To obtain the appropriate distribution, we compute the angular distribution of the current density from a QPC with width $W_i=7.1$~nm using Kwant.
The simulated device is depicted in Fig.~\ref{FigNumerics}c.
We then fit the resulting distribution as
\begin{equation}
\frac{dI}{d\theta}(\theta, \theta_0, \Gamma)= \frac{I}{N} \left[G(\theta, \theta_0, \Gamma)+G(\theta, -\theta_0, \Gamma)\right],
\label{EquationDist}
\end{equation}
where $G(\theta, \theta_0, \Gamma)=$ is a gaussian distribution, $\theta_0$ is the peak position, $\Gamma$ is the width, and $N=\int_{-\pi/2}^{\pi/2}(G(\theta, \theta_0, \Gamma)+G(\theta, -\theta_0, \Gamma)) d\theta$ the normalization factor. 
The fitted data is shown in Fig.~\ref{FigNumerics}d.
Since each gaussian corresponds to one of the valleys, we then use the corresponding sign of $theta_0$ for each Fermi surface in Eq.~\ref{FermiSurfaceGr}.
We also use a smaller width ($\Gamma / 8$) to obtain narrow TEF peaks as the ones observed in the measurements.

\subsection{Semiclassical calculations}

Using the Fermi surface obtained by fitting the tight-binding results with Eq.~\ref{FermiSurfaceGr} (Fig.~\ref{FigNumerics}b, red line), we compute the electron trajectories using $\vec{r}=(x,y)=\hbar(d\vec{k}_\tau/d\phi)/(e B)$ and assuming specular reflection~\cite{lee2016}.

To obtain the TEF spectra plotted in Figs.~2e and 2f of the main manuscript we assumed that the injector is a point contact with $W_i\ll L$ and the collector has a finite width $W_c$.
For every injection angle $-\pi/2<\theta<+\pi/2$, we compute the electron trajectory to determine whether it will reach the detector (that is, for all $\vec{r}$ where $y=0$, we determine if $L<x<L+W_c$).
We have assumed that the current dependency on $\theta$ follows the distribution shown in Fig.~\ref{FigNumerics}d.
Each trajectory that hits the collector adds $dI/d\theta$ to the collected current.
The final result at each $B$ is obtained by summing all the contributions in an equally-spaced distribution of $\theta$ between $-\pi/2$ and $\pi/2$ multiplied by the corresponding $dI/d\theta$.

\end{document}